\documentclass[usenatbib]{mnras}

\usepackage{newtxtext,newtxmath}
\usepackage[T1]{fontenc}
\usepackage{ae,aecompl}
\usepackage{docmute}
\usepackage{float}
\usepackage{comment}
\usepackage{color}
\usepackage{caption}
\usepackage{multirow}
\usepackage{docmute}
\usepackage{import}
\usepackage[dvipsnames]{xcolor}
\usepackage{graphicx}	
\usepackage{amsmath}	
\usepackage{amsfonts}
\usepackage{booktabs}
\usepackage{siunitx}
\usepackage[dvipsnames]{xcolor}
\usepackage{tikz,hyperref}
\usepackage{aas_macros}

\newcommand{\gizmo}{{\footnotesize GIZMO}}

\newcommand{\ahf}{{\footnotesize AHF}}

\hypersetup{
  linkcolor  = blue,
  citecolor  = blue,
  urlcolor   = magenta,
  colorlinks = true,
}

\title[Surface-density profiles of $\Lambda$CDM haloes]{An analytic surface density profile for \boldmath$\Lambda$CDM haloes and gravitational lensing studies}

\author[Lazar et al.]{Alexandres Lazar$^{1,2}$\thanks{\href{mailto:aalazar@uci.edu}{aalazar@uci.edu}},
James S. Bullock$^{1}$, 
Anna Nierenberg$^{3}$,
Leonidas A. Moustakas$^{2}$, and
\newauthor Michael Boylan-Kolchin$^{4}$
\\
$^{1}$Department of Physics and Astronomy, University of California, Irvine, CA 92697 USA\\
$^{2}$Jet Propulsion Laboratory, California Institute of Technology, 4800 Oak Grove Dr., Pasadena CA 91109, USA\\
$^{3}$Department of Physics, University of California Merced, 5200 North Lake Rd. Merced, CA 95343, USA\\
$^{4}$Department of Astronomy, The University of Texas at Austin, 2515 Speedway, Stop C1400, Austin, Texas 78712-1205, USA
}

\date{Working Draft\vspace{-0.6cm}}
\pubyear{2021}

\begin{document}
\label{firstpage}
\pagerange{\pageref{firstpage}--\pageref{lastpage}}
\maketitle

\begin{abstract}
We introduce an analytic surface density profile for dark matter haloes that accurately reproduces the structure of simulated haloes of mass $M_{\rm vir} = 10^{7-11}\ M_\odot$, making it useful for modeling line-of-sight perturbers in strong gravitational lensing models.  The two-parameter function has an analytic deflection potential and is more accurate than the projected Navarro, Frenk \& White (NFW) profile commonly adopted at this mass scale for perturbers, especially at the small radii of most relevant for lensing perturbations. Using a characteristic radius, $R_{-1}$, where the log slope of surface density is equal to $-1$, and an associated surface density, $\Sigma_{-1}$,  we can represent the expected lensing signal from line-of-sight halos statistically, for an ensemble of halo orientations, using a distribution of {\em projected concentration} parameters, $\mathcal{C}_{\rm vir} := r_{\rm vir}/ R_{-1}$. Though an individual halo can have a projected concentration that varies with orientation with respect to the observer, the range of projected concentrations correlates with the usual three-dimensional halo concentration in a way that enables ease of use. 
\end{abstract}

\begin{keywords}
cosmology:theory -- dark matter -- gravitational lensing: strong
\end{keywords}

\raggedbottom
\section{Introduction}

The $\Lambda$CDM (the cosmological constant + cold dark matter) cosmogony has served as the benchmark model for decades. A major component of its success is in matching the large-scale structure of the Universe, which also places principal constraints on the Universe's composition \citep[e.g.][]{Davis1985,Geller1989,Bond1996,Tegmark2004,Sanchez2006,Weinberg2013}. A key prediction of $\Lambda$CDM is that the Universe is lavished with a high number density of very low-mass ( $M_{\rm halo} \lesssim 10^{9}\ M_{\odot}$) dark matter halos that can act as perturbers in lensing studies \citep{Press1974,Metcalf2001,Green2005,Diemand2007,Springel2008,Frenk2012}. The existence of very low mass, nearly starless, dark halos of the kind and character predicted have yet to be confirmed by observations and may only be detectable via their gravitational effects e.g. on strong gravitational lenses.

While the cosmological model of $\Lambda$CDM has been successful at matching large-scale observations, $\Lambda$CDM still suffers from discrepancies found at small scales (see \citealt{Bullock2017} for a comprehensive overview) and this has motivated alternative dark matter models. One such model is warm dark matter (WDM), which suppresses the matter power spectrum of initial density perturbations at scales smaller than the free-streaming length \citep{Colin2000,Bode2001,Lovell2012,Schneider2012}. As an example, a 7~keV sterile neutrino of the type that could be responsible for the observed 3.5~keV line in galaxy cluster X-ray spectra \citep[e.g.][]{Boyarsky2014,Bulbul2014}, would produce a sharp cutoff in the abundance of halos smaller than $M_{\rm halo} \lesssim 10^{8}\ M_{\odot}$. Demonstrating the existence of halos below this mass could rule out this class of WDM; conversely, the ability to rule out such a population would eliminate $\Lambda$CDM as a complete model of cosmology. One other prediction for galaxies in a WDM cosmology is that they form with lower central densities compared to $\Lambda$CDM \citep{Lovell2014}. It is also possible that dark matter particles interact among themselves within a hidden sector \citep{Feng2009}. If self-interaction cross-sections are high enough, low-mass dark matter halos will have lower central densities compared to those of $\Lambda$CDM \citep{Rocha2013,Elbert2015,Tulin2018}. For dark matter halos of a given mass, all of these alternative cosmological paradigms make distinct predictions for both the abundance and central densities compared to expectations of $\Lambda$CDM.

Regardless of the cosmological model, a common characteristic of low-mass dark matter halos is that they are extremely inefficient at forming stars and are nearly devoid of baryons \citep[e.g.][]{Sawala2016,Benitez-Llambay2020}. This results in these objects being too difficult to detect electromagnetically. A more promising route for their detection is via gravitational perturbations of strongly lensed images \citep{Dalal2002,Moustakas2003,Koopmans2005,Vegetti2009a,Vegetti2009b,Vegetti2010,Vegetti2012,Hezaveh2016,Nierenberg2020,Gilman2020a,Hsueh2020}. To fully constrain the nature of dark matter using strong lensing, it is necessary to know the expected number of halos in $\Lambda$CDM and other cosmology models. It is convenient to characterize low-mass halo perturbers as either subhalos embedded within the main lens halo or field halos that are not part of the main lens but rather lie in projection near the lens' Einstein radius. In the strong lensing literature, this population of field halos are typically called ``line-of-sight'' (LOS) halos (or LOS perturbers). In \cite{Metcalf2005}, they showed that the LOS component contributes significantly to strongly lensed signals within $\Lambda$CDM. \cite{Li2017} later found that that LOS halos dominated the subhalo signal by a factor of 3-4. A similar result was found in \cite{Despali2018}, where the number of LOS perturbers compared to subhalo perturbers ranges from 3 to 10 times depending on the lens configuration. Similar results are found in \cite{Gilman2019}, \cite{He2021}, and \cite{Lazar2021}.

The internal structure of dark matter halo perturbers can greatly impact their lensing effect. Specifically, in order to properly constrain the mass function predicted by $\Lambda$CDM, the surface density profiles of the dark matter halos of interest must be precisely known \citep[e.g.,][]{Minor2017,Minor2021}. A common way to characterize low-mass dark matter halo structure for perturbing, field (LOS) halos is to assume that they follow spherically-symmetric NFW profiles \citep{Navarro1997,Nierenberg2017,Gilman2018,He2022}. 
With this assumption, the expected distribution of NFW concentration parameters at fixed mass becomes an important input for models \citep[e.g.][]{Gilman2022}. 

The spherical NFW profile is simpler analytically, which has an advantage that enables direct connection with theoretical predictions, especially for low-mass halos where feedback is believed to be less important in altering the density structure compared to dark-matter-only predictions \citep[e.g.][]{Lazar2020a}. Additionally, the NFW form also has a fairly easy-to-use surface-density profile for lensing-based analysis \citep{Wright2000}. However, there are some potential shortcomings in this approach. One is that dark matter halos are not perfectly spherical and tend to be more triaxial \citep[e.g.][]{Frenk1988,Cole1996,Allgood2006}. This means that, even for a halo with a spherically-averaged profile that is well described by an NFW fit, its surface density could well be different from the two-dimensional projection inferred by the spherical-average fit depending on its orientation. Second, it is well known that dark-matter-only simulations produce halos that are better modeled using the \cite{Einasto1965} profile than NFW when the particle resolution is increased \citep{Wang2020}. However, transforming that equation into the projected two-dimensional density and the lensing signal results in a relatively complicated expression \citep{Retana-Montenegro2012} which can be better approximated \citep{Dhar2010, Dhar2021}.

The aim of this work is to provide an easy-to-use surface-density profile that accurately reproduces the structure of simulated dark matter halos in any projection. Below we show with regression analysis that the projected density profiles of simulated dark matter halos deviate from the projected NFW profile inferred from their three-dimensional, spherical NFW fits (i.e., their best-fit fit concentrations and normalization) in ways that do not average out over all orientations. This is particularly true at projected radii smaller than the scale radius.  Since the NFW form is commonly used for substructure lensing analysis, this motivates the exploration of an alternative approach that is just as easy to use and more accurate. The projected profile presented below has a scale radius and a corresponding ``projected concentration.'' This projected concentration can account both for the fact that individual halos have different projected density profiles depending on their orientation with respect to the observer and that, at fixed halo mass, there is an intrinsic halo-to-halo (spherical) concentration scatter. While this paper focuses on circularly-averaged projected densities on the sky, the functional form can be easily generalized to elliptically-averaged profiles with one additional parameter, which would be a natural extension of this work. Note, however, that the corresponding extension for the lensing potential, reflection, and shear will be less straightforward \citep[see, e.g.][]{Tessore15,Riordan21}.

This paper is structured as follows. Section~\ref{sec:methods} introduces the suite of high-resolution simulations of dark matter halos used, provides a description of the selected sample of halos, and details the methods of constructing the radial dark matter profiles of each sampled halo. Section~\ref{sec:projected.structure} provides the main results of this paper and introduces our simple fitting formula for LOS perturbers for lensing analysis. We discuss the implications for our fitting formula in Section~\ref{sec:discussion} and demonstrate its impact to the commonly used LOS perturber formula, the NFW profile. Finally, we summarize out results in Section~\ref{sec:conclusion}.

\section{Methodology}
\label{sec:methods}
We use ``zoom-in'' {\em dark matter only} (DMO) simulations to study halo surface density structure at high resolution, focusing on isolated dark matter halos within the mass range that is applicable to substructure lensing, $M_{\rm halo} \in 10^{7-11}\, M_{\odot}$, at a characteristic redshift of $z = 0.2$. We use these simulations to discover an accurate surface density profile shape and to compare the shape parameters to the three-dimensional profile parameters of the same halos. We plan to follow-up with an analysis that utilizes a high-resolution cosmological environment (i.e., a cosmological box) to put the results presented here in a proper statistical setting and to explore redshift dependence. 

\subsection{Numerical zoom-in simulations}
All of our simulations use the multi-method code \gizmo{} \citep{Hopkins2015}. The initial conditions were calculated using {\small MUSIC} \citep{Hahn2013} at a redshift $z\approx 100$ following the methodology outlined in \citet{Onorbe2013}. This approach identifies a region spanning several virial radii around a main halo to resolve and produces a volume of uncontaminated halos around the main halo, which we use to study lower mass systems. 

 The first set of volumes we study come from the DMO versions of the following main halos from \cite{Fitts2017}, which assume a WMAP year 7 cosmology:  \texttt{m10f}, \texttt{m10g}, \texttt{m10h}, \texttt{m10i}, \texttt{m10j}, \texttt{m10k}, \texttt{m10l}, and \texttt{m10m}; these will be collectively referred to as the ``{\tt m10}'' suite of simulations. These simulations have a dark matter particle mass of $m_{\rm dm} = 3000\ M_{\odot}$ with a physical force resolution of $\epsilon_{\rm dm} = 35\ \rm pc$. At this resolution, we are able to also explore dark matter halos surrounding the main halo down to masses of $10^{7}\ M_{\odot}$, which contains ${\sim} 10^{4}$ particles within the virial radius. In addition, we use several of the dark matter only volumes surrounding  main halos mass of $\sim 10^{11}\, M_{\odot}$, which were first presented in \cite{Lazar2020a}: \texttt{m11d}, \texttt{m11e}, \texttt{m11h}, and \texttt{m11i}; these will be collectively referred to as the ``{\tt m11}'' suite of simulations. The \texttt{m11} simulations have a mass resolution $\sim$~14 times coarser than the \texttt{m10} suite. Within the {\tt m11} volumes, we only explore the main halo with $M_{\rm vir} \sim 10^{11}\, M_{\odot}$. These simulations use a Planck 2015 cosmology \citep{PlanckCollaboration2016}. 

Dark matter halos in all the simulations considered here are identified using the Amiga halo finder (\ahf; \citealt{Knollmann2009}). The halo finder uses a recursively refined grid that determines the local overdensities found within the density field and identifies the density peaks of this field as the center of these halos.

\subsection{Dark matter halo nomenclature}
In this section we provide definitions of the global and physical quantities of dark matter halos we will be studying in this work. Throughout this paper, lower case $r$ denotes the physical, de-projected, three-dimensional radius. The projected, two-dimensional radius is denoted by an upper case $R$ without a subscript.

\subsubsection{Halo mass and radius}
Throughout this paper, dark matter halos are defined to be spherical systems with virial mass, $M_{\rm vir}$, with a virial radius, $r_{\rm vir}$, inside of which the average density is equal to the critical density times a multiplicative factor $\Delta_{\rm vir}(z)$,  i.e.,
\begin{align}
    M_{\rm vir}
    = \frac{4\pi}{3} r_{\rm vir}^{3}\, \Delta_{\rm vir}(z)\, \rho_{\rm c}(z)
    \, .
\end{align}
Here, $\Delta_{\rm vir}$ is redshift-dependent virial overdensity \citep{Bryan1998} and $\rho_{\rm c}(z) = 3 H^{2}(z)/8\pi G$ is the critical density of the universe at redshift $z$. The \cite{Bryan1998} definition is the primary spherical overdensity definition used for the \ahf{} halo finding, but is computed using the gravitationally bound particles of the system. For the purposes of this analysis, $M_{\rm vir}$ and $r_{\rm vir}$ are recomputed using {\em all} dark matter particles (bound and unbound) of the spherical overdensity peak identified by \ahf. It should be noted that isolated dark matter halos are also commonly defined with overdensity $\Delta = 200$ with either a critical or mean-matter background densities. We do not explore such definitions other than $M_{\rm vir}$ in this analysis, but will do so in a follow up paper.

\subsubsection{Three-dimensional structure of dark matter halos}
The three-dimensional density profiles of dark matter halos in $\Lambda$CDM are commonly modeled with the \cite*{Navarro1997} (NFW) profile,
\begin{align}
    \rho_{\rm NFW}(r) = \rho_{s} \left( \frac{r}{r_{s}} \right)^{-1} \left( 1 + \frac{r}{r_{s}} \right)^{-2}
    \, ,
    \label{eq:density.nfw}
\end{align}
where $r_{s}$ is the scale radius and $\rho_{s}$ is the characteristic density. The structure of a NFW halo is parameterized by the concentration ($c_{\rm NFW}$) of a dark matter halo, which is formally defined as the ratio between the size of the halo, $r_{\rm vir}$, and the scale radius, $r_{s}$: 
\begin{align}
    c_{\rm NFW} := r_{\rm vir} / r_{s}
    \, .
    \label{eq:conc.nfw}
\end{align}

In DMO simulations, dark matter halos are better described by the Einasto profile \citep{Navarro2004,Navarro2010}, which provides a good description over 20 decades of halo masses \citep{Wang2020}:
\begin{align}
    \rho_{\epsilon}(r) 
    &=
    \rho_{-2} \exp\Bigg\{
    -\frac{2}{\alpha_{\epsilon}}
    \Bigg[ \Bigg(\frac{r}{r_{-2}}\Bigg)^{\alpha_{\epsilon}}-1 \Bigg]
    \Bigg\}
    \label{eq:density.einasto}
    \, .
\end{align}
Here, $r_{-2}$ is the scale radius where the log-slope is equal to $-2$, the scale density is $\rho_{-2} := \rho(r_{-2})$, and $\alpha_{\epsilon}$ is the shape parameter. Usually $\alpha_{\epsilon}$ is fixed to make this a two-parameter function, but it can be expressed in terms of peak height \citep[e.g.][]{Gao2008,Prada2012,Klypin2016,Child2018}. We fix $\alpha_{\epsilon} = 0.17$ for this analysis, which fits well for the range of halo masses we explore here and is close to commonly-adopted values \citep[e.g.][]{Wang2020}. The internal structure from the Einasto profile is parameterized by the dimensionless concentration parameter:
\begin{align}
    c_{\rm vir} := r_{\rm vir}/r_{-2}
    \, .
    \label{eq:conc.einasto}
\end{align}
Note that $r_s$ in the NFW profile is also the radius where the log slope of the NFW form is $-2$, so typically $c_{\rm vir} \simeq c_{\rm NFW}$ for any individual halo.

\subsubsection{Two-dimensional structure of dark matter halos}
Dark matter halos in substructure lensing are modeled as two-dimensional surface density profiles, as these coincide with two-dimensional, face-on projections of observations viewed from the surface of the sky. For a given system that has a (intrinsic) spherically-averaged local density profile in three-dimensions, $\rho(r)$, the {\em cylindricaly-averaged, local surface density} profile, $\Sigma_{\rho}(R)$, is quantified by integrating the along the line of sight, $\ell$, for a path of some depth $\mathcal{L}$
\begin{align}
    \Sigma_{\rho}(R) 
    &= \int^{\mathcal{L}/2}_{-\mathcal{L}/2} d\ell\, \rho(R, \ell) \label{eq:integrated.projection} \\
    &= 2 \int^{\mathcal{L}/2}_{R} dr\, \frac{r \rho(r)}{\sqrt{r^{2} - R^{2}}}
    \, ,
\end{align}
where $R$ denotes the projected radius relative to the center of the halo and $\ell = \sqrt{r^{2}-R^{2}}$ is a coordinate along the line of sight to the observer. In what follows we explore choices for the path length of integration in constructing numerical surface density profiles of the form $\mathcal{L} = 2\, \xi\, r_{\rm vir}$.  As described below, we settle on $\xi = 1.5$ as an optimal choice.

In lensing studies, it is common to use an NFW profile in the limit where $\mathcal{L} = \infty$ via a forward Abel transformation. In this case, the surface density profile is analytic \citep{Wright2000}:
\begin{align}
    \Sigma_{\rm NFW}(x) = 
    \begin{cases} 
    \ \frac{2\, \rho_{s} r_{s}}{(x^{2} - 1)} \left[ 1 - \frac{2}{\sqrt{x^{2} - 1}} \arctan\sqrt{ \frac{x - 1}{1 + x} } \right] & x > 1 \\
    \ \frac{2\, \rho_{s} r_{s}}{3} & x = 1 \\
    \ \frac{2\, \rho_{s} r_{s}}{(x^{2} - 1)} \left[ 1 - \frac{2}{\sqrt{1 - x^{2}}} \mathrm{arctanh}\sqrt{ \frac{1 - x}{1 + x} } \right] & x < 1 
   \end{cases}
   \, ,
   \label{eq:pdensity.nfw}
\end{align}
with $x = R/r_{s}$. While the Einasto profile describes the spherically averaged distribution of high-resolution dark matter halos to better accuracy than NFW (see Figure~\ref{fig:density.compare} below), the analytical lens properties are significantly more complicated to work with \citep[see][]{Retana-Montenegro2012}. For this reason, the NFW profile is most commonly adopted in lensing studies. In later sections we present an analytic surface density profile that describes projected dark matter halo structure with a similar accuracy as the Einasto profile does in three-dimensions, with the added feature that it has an easy-to-use, analytic lensing potential.

\subsubsection{Non-spherical dark matter halos}
\label{sec:halo.shape}
Another important aspect of dark matter halos is their non-spherical shape. $\Lambda$CDM halos are triaxial and tend to be more elongated at higher halo masses. Dark matter halos viewed along their major axis (i.e., the densest axis) could be misidentified as a higher mass halo or a halo of higher concentration. To explore the magnitude of this effect, we calculate dark matter halo shapes for our halos by computing the shape inertia tensor outlined in \cite{Allgood2006}. This is done by solving the eigenvalues from all of the dark matter particles within a shell between 10 - 20\% of $r_{\rm vir}$. The resulting eigenvalues of the shape tensor are proportional to the square root of the principal axes of the dark matter distribution, which we will refer to as the ``major'', ``intermediate'', and ``minor'' axes throughout this paper. 

Note that the triaxial shape of DMO halos can be taken into account in lensing studies by projecting a triaxial NFW profile directly \citep[e.g.,][]{Feroz12}. Doing so requires six parameters rather than two for each halo. In what follows, we present a direct fit to the projected, cylindricaly-averaged, profiles that provide improved accuracy over the projected NFW with the same number of free parameters.

\begin{figure*}
    \centering
    \includegraphics[width=\textwidth]{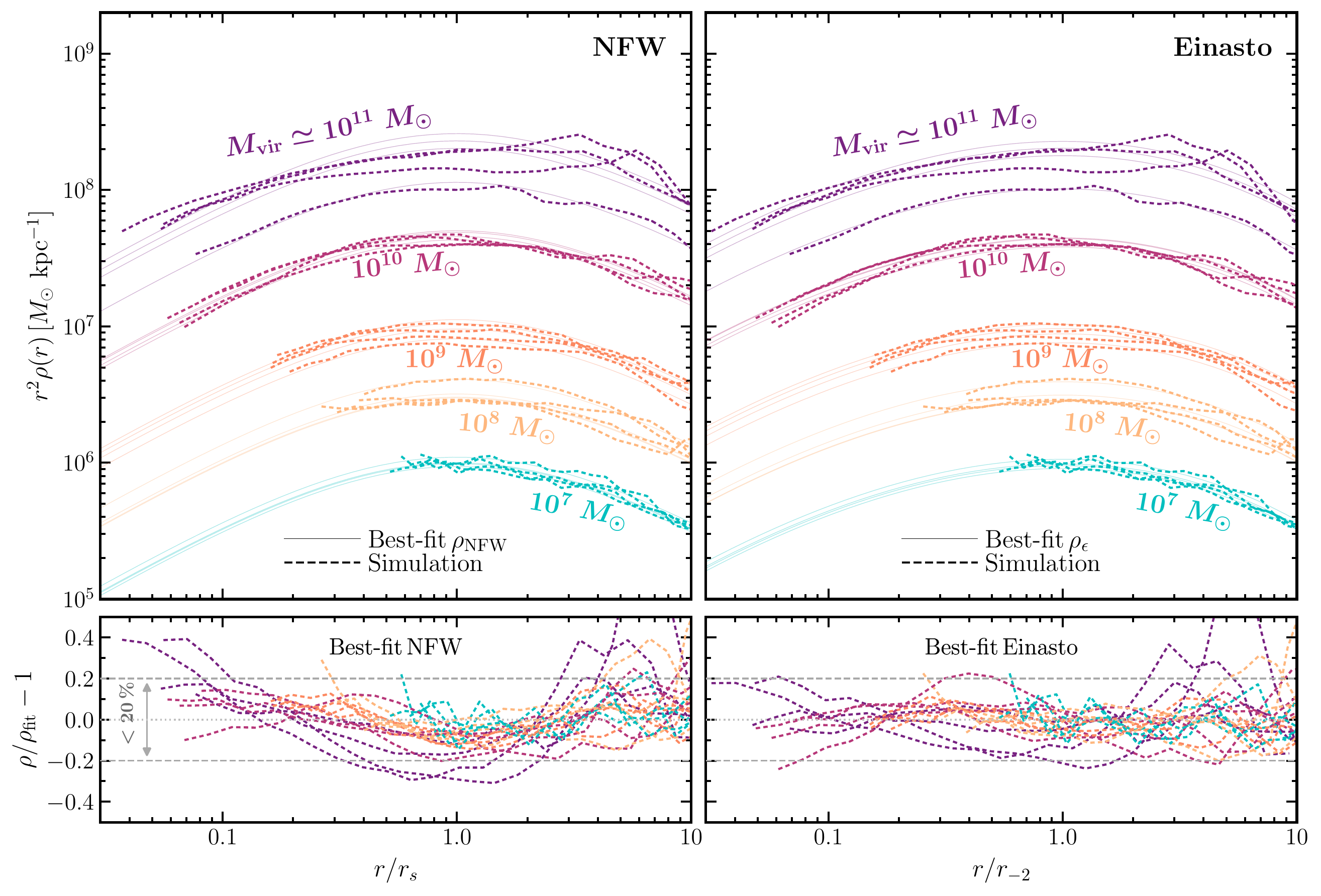}
    \caption{
        The spherically averaged density profiles, $\rho(r)$, of the simulated dark matter halos at $z = 0.2$. The top panels present several density profiles of the isolated dark matter halos (dashed curves) in mass groups spanning $M_{\rm vir} = 10^{7-11}\ M_{\odot}$. The density profiles are scaled by $r^{2}$ to set apart the dynamic range and the $x$-axis is scaled by $r_{s}$ of the best-fit NFW profile (left panels) and $r_{-2}$ of the best-fit Einasto profile (right panels). The simulated profiles are plotted from the convergence radius, $r_{\rm conv}$ out to $r_{\rm vir}$. The thin solid lines illustrates the best-fit halo density profile of each halo for the NFW profile (left panel) and the Einasto profile (right panel). The bottom panels depicts the quality of fit based on the {\em best-fit} parameters for each individual halo using the NFW and Einasto density profiles. The gray dashed lines encapsulates fits within 20\% accuracy.
    }
    \label{fig:density.compare}
\end{figure*}

\subsection{Constructing and fitting radial profiles}

\subsubsection{Region of numerical convergence}
One of the key components of our analysis focuses on the spherically (and cylindrically) averaged mass-density profiles. Before radial bins are constructed, dark matter particles are first shifted relative to the halo center determined by \ahf. The innermost regions of $N$-body simulated dark matter halos, to some extent, are impacted by numerical relaxation. The region of convergence, $r_{\rm conv}$, is quantified using the method specified in \cite{Power2003}, where the effective resolution of the simulations dictates the location of the radius where the two-body relaxation timescale, $t_{\rm relax}$, becomes shorter than the age of the universe, $t_{0}$, set by the criterion:
\begin{align}
    \frac{t_{\rm relax}(r)}{t_{0}} = \frac{\sqrt{200}}{8}\frac{N(<r)}{\ln N(<r)}
    \left[ \frac{\bar{\rho}(<r)}{\rho_{\rm c}(z)} \right]^{-1/2}
    \, .
    \label{eq:power.radius}
\end{align}
Here, $N(<r)$ is the cumulative number of particles within radius $r$ and the cumulative density profile, $\bar{\rho}(<r) = 3M (<r)/4\pi r^{3}$, with $M(<r)$ being the cumulative mass. For $N$-body simulations of our resolution, convergence is shown to be well resolved to the radius at which the criterion satisfies $t_{\rm relax} >0.6 \, t_{0}$ with <1\%  deviations for isolated zoom runs (see \citealt{Hopkins2018}). The typical convergence radii within our sample at redshift $z=0.2$ are resolved to regions relevant for lensing based analysis; in what follows, we present results grouped in five mass bins, set by five decades in halo virial mass: $M_{\rm vir}\simeq 10^{7-11}\, M_{\odot}$. The halo masses with bins of $M_{\rm vir} = \{10^{7},\, 10^{8},\, 10^{9},\, 10^{10},\, 10^{11}\}\ M_{\odot}$ have convergence radii of $r_{\rm conv} \simeq \{300,\, 270,\, 240,\, 200,\, 500\}\ \rm pc$, in order from lowest to highest mass decade.

\begin{figure*}
    \centering
    \includegraphics[width=\textwidth]{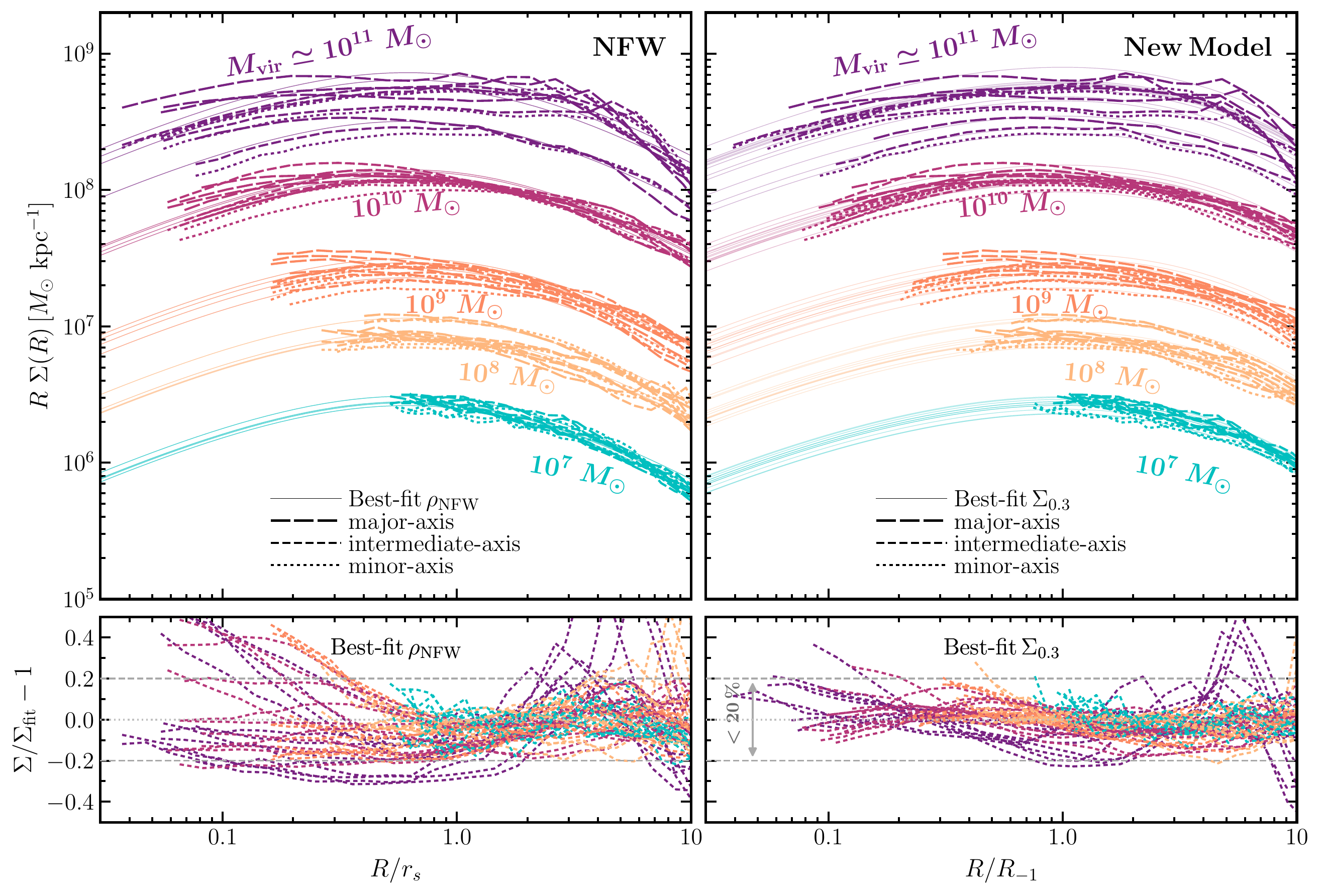}
    \caption{
        The resulting {\em mean surface density} profiles of the simulated dark matter halos at $z = 0.2$. Similar in presentation as the previous figure, the top panel presents the same isolated dark matter halos, but now for their projected density profiles, $\Sigma(R)$, along the major (thick-dashed curve), intermediate (thin-dashed curve), and minor axis (dotted curve). The bottom panels depict the fit quality made to each density axis. Here, we scale the $\Sigma(R)$ with $R$ to emphasize the dynamic range and we also scale the projected radius ($R$) by the respective, best-fit scale radii, where the simulated profiles in $R$ are plotted from the convergence radius, $r_{\rm conv}$ out to $r_{\rm vir}$. {\em Left:} The thin solid line shows the expected surface-density profile of each halo from the previous figure according to the NFW model, which are based on their {\em best-fit} parameters for each individual three-dimensional density profile. The radii is also scaled by the corresponding scale radius, $r_{s}$). {\em Right:} The same simulated curves presented in the left panel, but instead, the {\em best-fit} curves from fitting the $\Sigma_{0.3}$ profile (Equation~\ref{eq:general.proj.profile} with $\beta = 0.3$) we introduced in the main text applied to each density projection. The projected radii for these panels are normalized by the radius at which the logarithmic slope is equal to $-1$, $R_{-1}$, which is acquired by the curve fitting procedure. The $\Sigma_{0.3}$ profile is able to model each orientation to $\sim 20$ \% and better based off of the fit quality presented in the bottom panel; an improvement from the NFW model shown in the left panel. 
    }
    \label{fig:surface.density.compare}
\end{figure*}

\subsubsection{Constructing spherical density profiles}
For each halo, we construct density profiles, $\rho(r)$, using the {\em total} (bound and unbound) dark matter mass found in radial shells divided by the volume of each shell. We fit NFW and Einasto profiles to each profile using 35 logarithmic-spaced radial bins from $r_{\rm conv}$ to $r_{\rm vir}$.  The best-fit parameters are determined by the following minimization of the figure-of-merit in a least-squares optimization:\footnote{The convention used here in this paper is $\log \equiv \log_{e} \equiv \ln$.}
\begin{align}
    Q_{\rho}^{2} 
    = 
    \frac{1}{N_{\rm bins}} \sum_{i}^{N_{\rm bins}}
    \left[
    \log\left(r_{i}^{2}\rho_{i}\right) - \log\left(r_{i}^{2}\rho_{i}^{\rm model}\right)
    \right]^{2}
    \, .
    \label{eq:merit.density}
\end{align}
Instead of $\rho$ being minimized, which has the numerical value decrease by many orders of magnitude between the inner and outer region of the profile, the $r^{2}\rho$ merit provides a more balanced indicator of goodness-of-fit across the entire radial range.\footnote{The choice of a scaled density as the function to be minimized for the curve fitting, has been more often used (e.g. \citealt{Navarro2004}), since the scale density fits the entirety of the density profile without weighing the inner-density more, while a density ($\rho$) minimization would bias this. Both minimization definitions have been compared quantitatively, and we find the scaled minimization performs better.} From each individual fit, we record best-fit parameters for the NFW density profile and Einasto profile with fixed $\alpha_{\epsilon} = 0.17$. 

\subsubsection{Constructing cylindrical surface-density profiles}

Halo surface density profiles are constructed in a similar manner to the three-dimensional profile: we measure the local surface density profile in two-dimensions, $\Sigma(R)$, by counting {\em all} (bound and unbound) dark matter particle mass that exists along a line-of-sight of total depth $\mathcal{L}$ in a thin circular ring of average radius $R$ divided by the area of that ring.  We use 35 logarithmic spaced bins of projected radius $R$ that span an inner radius $R = r_{\rm conv}$ to an outer cylindrical radius $R = r_{\rm vir}$. 

In order to define the surface density of a three-dimensional object, we must define a depth of interest, $\mathcal{L}$ (see Equation \ref{eq:integrated.projection}). In analytic investigations, it is common to chose $\mathcal{L} = \infty$; however this is neither possible numerically nor physically meaningful since we would like to characterize the degree of perturbation caused by the halo {\em over the background}. At minimum, one should span the full diameter of the halo and use  $\mathcal{L} = 2\, r_{\rm vir}$. However, we expect that there is additional clustered matter outside $r_{\rm vir}$. We have explored this question using a depth parameter $\xi$ defined via $\mathcal{L} = 2\, \xi\, r_{\rm vir}$. As discussed in  Appendix~\ref{sec:projection.depth}, we find that $\xi = 1.5$ provides stable results that appear to capture the relevant over-density adequately. We use $\xi = 1.5$ for the rest of this work.

In the following subsections we discuss fits to projected profiles.  We do so  by minimizing the figure-of-merit in a least squares optimization for surface density structures in a given projection:
\begin{align}
    Q_{\Sigma}^{2} 
    = 
    \frac{1}{N_{\rm bins}} \sum_{i}^{N_{\rm bins}}
    \left[
    \log\left(R_{i}\Sigma_{i}\right) - \log\left(R_{i}\Sigma_{i}^{\rm model}\right)
    \right]^{2}
    \, .
    \label{eq:merit.proj.density}
\end{align}
As with the figure of merit used for the three-dimensional density profile, the $R\Sigma$ merit enables a more balanced weighing across the radial range being fitted. We exclude the outer radii $(R > 0.5r_{\rm vir})$ when fitting since we aim to model the inner, higher-density regions that are most relevant for strong-gravitational lensing studies. 

\section{The spherical and projected structure of CDM haloes}
\label{sec:projected.structure}
This section present the main results of this paper. We first start by contextualizing our work by comparing the accuracy of NFW and Einasto fits to the spherically-averaged density profiles for our sample of haloes (Section~\ref{sec:proj.surf.dens}). We then  present the circularly-averaged surface density profiles of the same dark matter halos  and introduce an easy-to-use analytic profile that improves upon the projected NFW case, with an accuracy similar to that of an Einasto in three dimensions (Section~\ref{sec:improved.model}). Finally, we introduce a projected concentration parameter as a convenient characterization of the surface density structure of haloes in Section~\ref{sec:param.orient}.

\subsection{Accuracy of NFW and Einasto profiles to simulations}
\label{sec:proj.surf.dens}
The top panels of Figure~\ref{fig:density.compare} plot the local, spherically-averaged density profiles, $\rho(r)$, for several simulated halos (colored dash curves) grouped by  decades of mass at $z = 0.2$. We chose this redshift since it is typical of lens systems.  The density is additionally scaled by $r^{2}$ to constrain the dynamic range of the horizontal axis. The halos with masses ranging from $10^{7} - 10^{10}\ M_{\odot}$ are collected from the \texttt{m10} suite of simulations while the \texttt{m11} simulations only presents their $10^{11}\ M_{\odot}$ main halos. Each profile is plotted from the virial radius, $r_{\rm vir}$, down to the innermost converged radius, $r_{\rm conv}$.\footnote{We adopt this convention for most of the figures in this paper, unless otherwise specified.} The thin solid lines show the best-fit NFW profiles (left plot) and Einasto profiles (right plot). The radii along the horizontal axis are all scaled by their best-fit scale radius, $r_{s}$ for the NFW profile fits and $r_{-2}$ for the Einasto profile fits.  Residuals of the fits are shown in the bottom panels. We see that best-fit NFW profiles capture the normalization of the simulated profiles to within $\sim 40\%$, though the residuals do show a systematic U-shape.  The best-fit Einasto profiles show no systematic difference with radius, and are accurate to within $\sim 20\%$ except at large radius. It is this level of improvement that has motivated the community to utilize the Einasto profile in favor of the NFW profile when precise predictions are required.  We present this comparison here to provide context for the new surface-density fit we propose. We note that the projected version of the Einasto profile is analytically cumbersome, and this is why most lensing searches for low-mass perturbers assume projected NFW forms.

The left panels of Figure~\ref{fig:surface.density.compare} presents the same simulated halos shown in Figure~\ref{fig:density.compare}, but now as circularly-averaged surface-density profiles, $\Sigma(R)$, along with the projected version of the best-fit NFW profile for each halo (Equation~\ref{eq:pdensity.nfw}, thin solid lines). For each halo, we show the surface density profile along the three main halo density axes (as described in Section~\ref{sec:halo.shape}): the major axis (thick-dashed curves), intermediate axis (thin-dashed curves), and the minor axis (dotted curves). On the vertical axis we plot the surface density profile multiplied by the projected radius, $R$, to limit the dynamic range. The horizontal axis shows the projected radius, $R$, divided by the best-fit NFW scale radius determine from the spherically average density fits (left plot), $r_{s}$.  The fit residuals are shown along the bottom.  We see that, especially at small radii, the fits are systematically biased (either low or high depending on the halo orientation).

In the next section, we present a profile, $\Sigma_{0.3}$ that captures the surface density profiles of cylindrical-averaged halos more accurately than the best-fit NFW average density profile.   

\subsection{An improved surface density profile for dark matter halos}
\label{sec:improved.model}


\begin{figure}
    \includegraphics[width=\columnwidth]{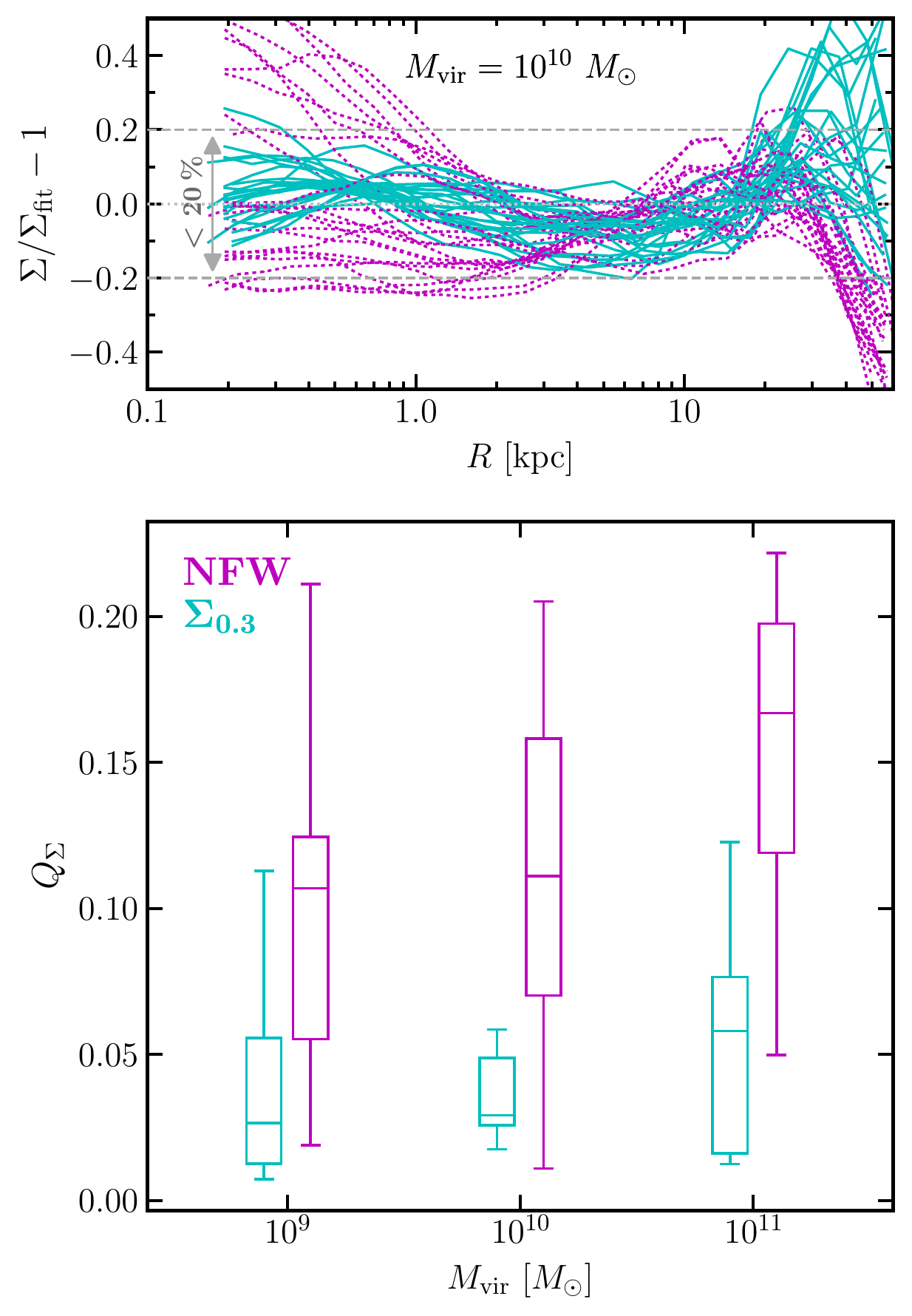}
    \caption{
        {\em Top}: The profile residuals for the eight $10^{10}\ M_{\odot}$ halos viewed along the major, minor, and intermediate density axes the NFW model (dashed magenta curves) and our $\Sigma_{0.3}$ model. Residuals for the NFW fit often exceed $20\%$ while the new treatment is significantly better, especially at small radii . {\em Bottom}: The distribution of figure-of-merits (Equation~\ref{eq:merit.proj.density}) for each halo mass bin for the two fits.  Here we evaluate the figure-of-merit using only bins with $R < R_{-1}$ in order to focus on the high-density region of most interest for lensing studies.
    }
    \label{fig:fit.performance}
\end{figure}

In Figure~\ref{fig:surface.density.compare} it is interesting to note that there is a near-constant turnover in the profiles as a function of projected radius.  It is instructive then to see how the logarithmic slope of the projected density, $\mathrm{d}\log{\Sigma} / \mathrm{d}\log{R}$, behaves a function of projected radius, $R$. We go into a much further discussion in Appendix~\ref{sec:log.slopes}, where we find that the behavior of the logarithmic slope for the surface-density profile is captured by a radially-dependent power law profile that is simple in form:
\begin{align}
    \frac{\mathrm{d}\log{\Sigma_{\beta}}}{\mathrm{d}\log{R}}
    =
    - \Bigg( \frac{R}{R_{-1}} \Bigg)^{\beta} 
    \, .
    \label{eq:proj.profile:slope}
\end{align}
Here, $R_{-1}$ is the projected radius where the logarithmic slope of the surface density is equal to $-1$ and $\beta$ is a shape parameter. For a fixed shape parameter, $\beta$,  a given orientation can be parameterized by the projected scale radius, $R_{-1}$, which in-turn captures the variation and scatter expected from halo-to-halo. Figure~\ref{fig:log.slope} shows that this profile, with $\beta$ fixed at $0.3$, reproduces the observed behavior of our simulated halos along multiple axes quite accurately, and is consistent with a lack of convergence to a central power law. 

Integrating Equation~\ref{eq:proj.profile:slope} gives us a generalized surface density profile:
\begin{align}
    \Sigma_{\beta}(R)
    =
    \Sigma_{-1}
    \exp\Bigg\{
    -\frac{1}{\beta}
    \Bigg[ \Bigg( \frac{R}{R_{-1}} \Bigg)^{\beta} - 1 \Bigg] \Bigg\}
   \label{eq:general.proj.profile}
   \, ,
\end{align}
where the normalization is defined by $\Sigma_{-1} := \Sigma(R_{-1})$. This has the same functional form as the \cite{Sersic1968} and \cite{Einasto1965} profiles. Note that at $R=0$ there is a finite surface density: $\Sigma(0) = \Sigma_{-1}\exp(1/\beta)$. As written, $\Sigma_{\beta}$ has three free parameters: $\Sigma_{-1}$, $R_{-1}$, and $\beta$. Leaving all three parameters free to fit provides a means to describe halo density structure extremely accurately.  Equation \ref{eq:mass1} in the appendix provides an analytic expression for the the mass within given projected radius $R$ for given these three parameters. Equation \ref{eq:mass.fit} provides an easy-to-use fitting function that relates the three parameters to the total projected mass within the virial radius (approximately equal to the three-dimensional virial mass, see Appendix~\ref{sec:projection.depth}).

Usefully, for the mass ranges of $\Lambda$CDM halos explored here, we find that $\beta=0.3$ allows us to characterize our entire sample of halos with two free parameters quite well, as shown below. With $\beta$ fixed to $0.3$,  we have a two-parameter function, $\Sigma_{0.3}(R)$, that can be fully specified by $\Sigma_{-1}$ and $R_{-1}$.  As demonstrated in Section~\ref{sec:param.orient}, we can also characterize the profile with a concentration parameter, in analogy with what is traditionally done with NFW profiles or Einasto profiles in three-dimensions. 

\begin{figure}
    \includegraphics[width=\columnwidth]{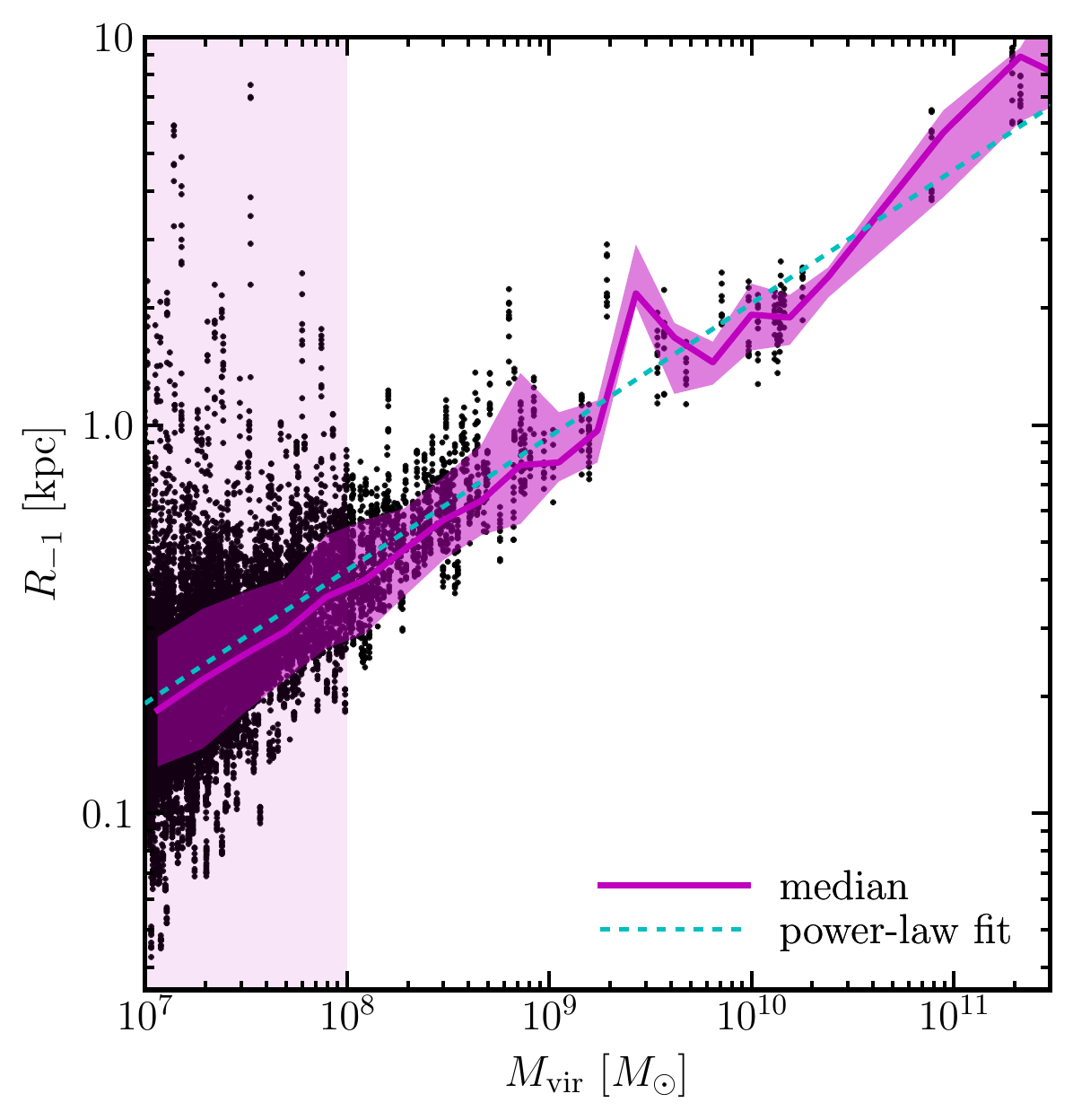}
    \caption{
        Best-fit values of the $R_{-1}$ scale radius versus halo virial mass for $\Sigma_{0.3}$ fits to 10 random projections to each halo in our sample. The solid line and shaded bands reflect the median and ninety percentile bands of the distribution.  The vertical pink band shows the region where fits are likely not reliable because the typical value of $R_{-1}$ is smaller than the typical convergence radius for halos. Dotted power-law fit shows the relation $R_{-1} \propto M_{\rm vir}^{0.35}$. Note that these simulations are not a representative cosmological sample, so this masss-radius relation may not be representative for the halo population in general.
    }
    \label{fig:R1-vs-Mvir}
\end{figure}

\begin{figure}
    \includegraphics[width=\columnwidth]{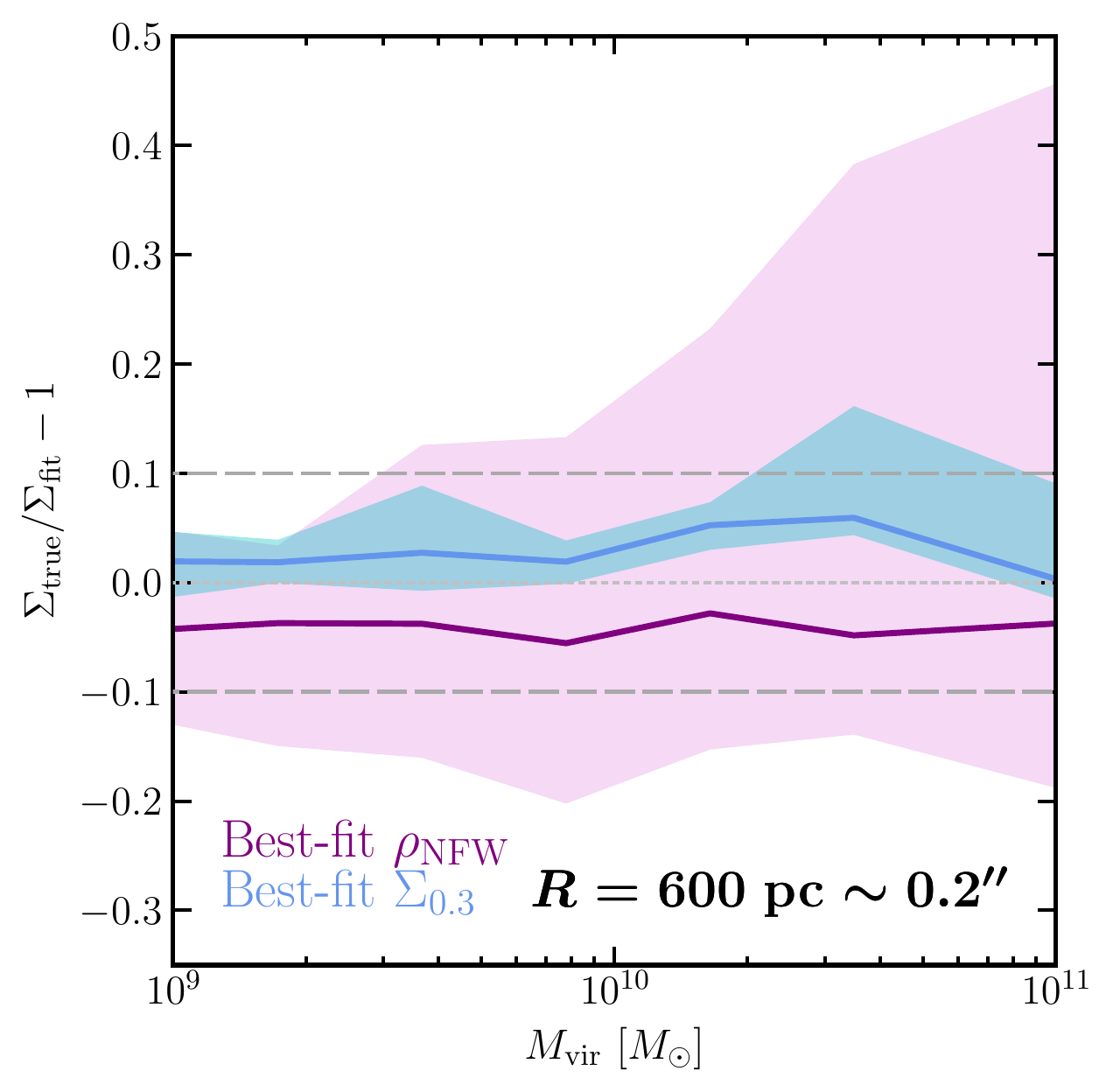}
    \caption{
    The quality of fit for inner-most projected density (at 600 pc) between the simulation results, $\Sigma_{\rm true}$, and the profile fits, $\Sigma_{\rm fit}$, for the $\rho_{\rm NFW}$ and $\Sigma_{0.3}$ profiles (purple and blue curves, respectively), as a function of $M_{\rm vir}$. The bands encapsulate the $2\sigma$ scatter for ten random projections on each resolved isolated halo. The relative error in NFW fits is larger at larger masses because $600$ pc is ``deeper" within the core (smaller compared to $R_{-1}$) at higher masses (see Figure \ref{fig:R1-vs-Mvir}).  The $\Sigma_{0.3}$ fits provide the most improvement for radii smaller than the scale radius $R < R_{-1}$.   
    }
    \label{fig:fit.quality-innerDensity}
\end{figure}

The quality of the $\Sigma_{0.3}$ fits can be seen in the right-hand panels of Figure~\ref{fig:surface.density.compare}.  Here we use the same simulated halos presented earlier in the left-most panel, but instead of using NFW fits for each halo, we now use the best-fit $\Sigma_{0.3}$ fit for the {\em individual} density-axis projections (the solid-thin lines). The projected radius along the horizontal axis is normalized by the best-fit projected scale radius, $R_{-1}$. The residuals of the $\Sigma_{0.3}$ fits are shown in the lower-right panel, where the inner-most region is captured to better than $\sim 20$\% in all cases. This a clear improvement to the NFW model shown on the left, and similar in quality to the Einasto profile fits of halos within three dimensions shown in Figure \ref{fig:density.compare}. The improvement with respect to NFW is most significant at small radii $R \lesssim R_{-1}$, where the surface densities are the highest.

To further emphasize the improvement of the $\Sigma_{0.3}$ fits compared to NFW, we present quality-of-fit parameters in Figure~\ref{fig:fit.performance}. The top panel shows residuals for the eight $10^{10}\ M_{\odot}$ halos in our sample for the NFW model (dashed magenta curves) and our $\Sigma_{0.3}$ model (solid cyan curves).  We use the projected profiles for each halo projected along its major, minor, intermediate density axes. We see that the NFW model can be off by as much as $\sim 50\%$ at small radius, and that the sign of the offset is systematic, depending on orientation.  The new model captures the profiles to within 20\% in all cases. At small radii, $R \lesssim 2$ kpc, which corresponds to $R \lesssim R_{-1}$ in this mass range, we again see that the $\Sigma_{0.3}$ profile is able to capture the true projected density structure more accurately than the inferred NFW shape.

Figure~\ref{fig:R1-vs-Mvir} shows the relationship between best-fit $R_{-1}$ values and halo $M_{\rm vir}$ for our full sample. We  have fit $\Sigma_{0.3}$ profiles to ten random projections of each halo.   The thick magenta line and associated band show the median and 90 percentile range as a function of virial mass. The dotted cyan line shows a power-law fit, which has a best-fit slope $R_{-1} \propto M_{\rm vir}^{\gamma}$, where $\gamma = 0.35$,  from only the sample up to $M_{\rm vir} = 10^{10}\ M_{\odot}$. The vertical pink band shows the region ($M_{\rm vir} < 10^8\, M_{\odot}$) where typical values of $R_{-1}$ are smaller than the typical convergence radii for the simulated halos. Fits in this region rely on extrapolation to infer the value of $R_{-1}$ and this likely gives rise to some non-physical dispersion in characteristic radius.

\begin{figure}
    \includegraphics[width=\columnwidth]{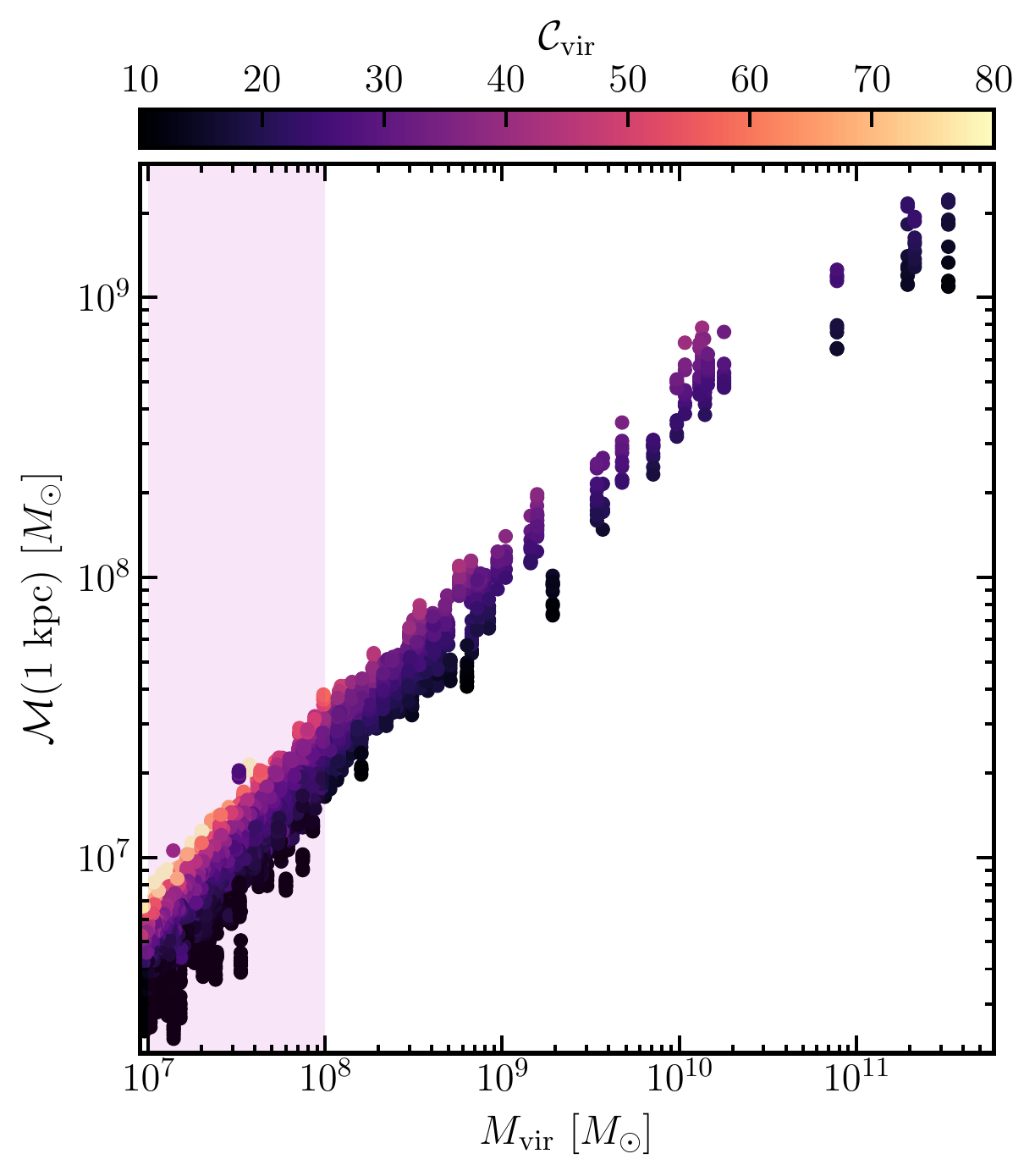}
    \caption{
        The projected mass within 1 kpc, $\mathcal{M}(<R = 1 {\rm kpc})$,  as a function of virial mass, $M_{\rm vir}$, for our sample of halos, each viewed along 10 random orientations.  The points are color coded by the projected concentrations associated with $\Sigma_{0.3}$ fits along each orientation. We see that the projected concentration correlates with the density of the halo along a given projection at fixed $M_{\rm vir}$, and that this accounts for as much as a factor of $\sim 2$ variation in the relationship between virial mass and projected mass at this radius. 
    }
    \label{fig:m1kpc-ceff}
\end{figure}

\begin{figure}
    \includegraphics[width=\columnwidth]{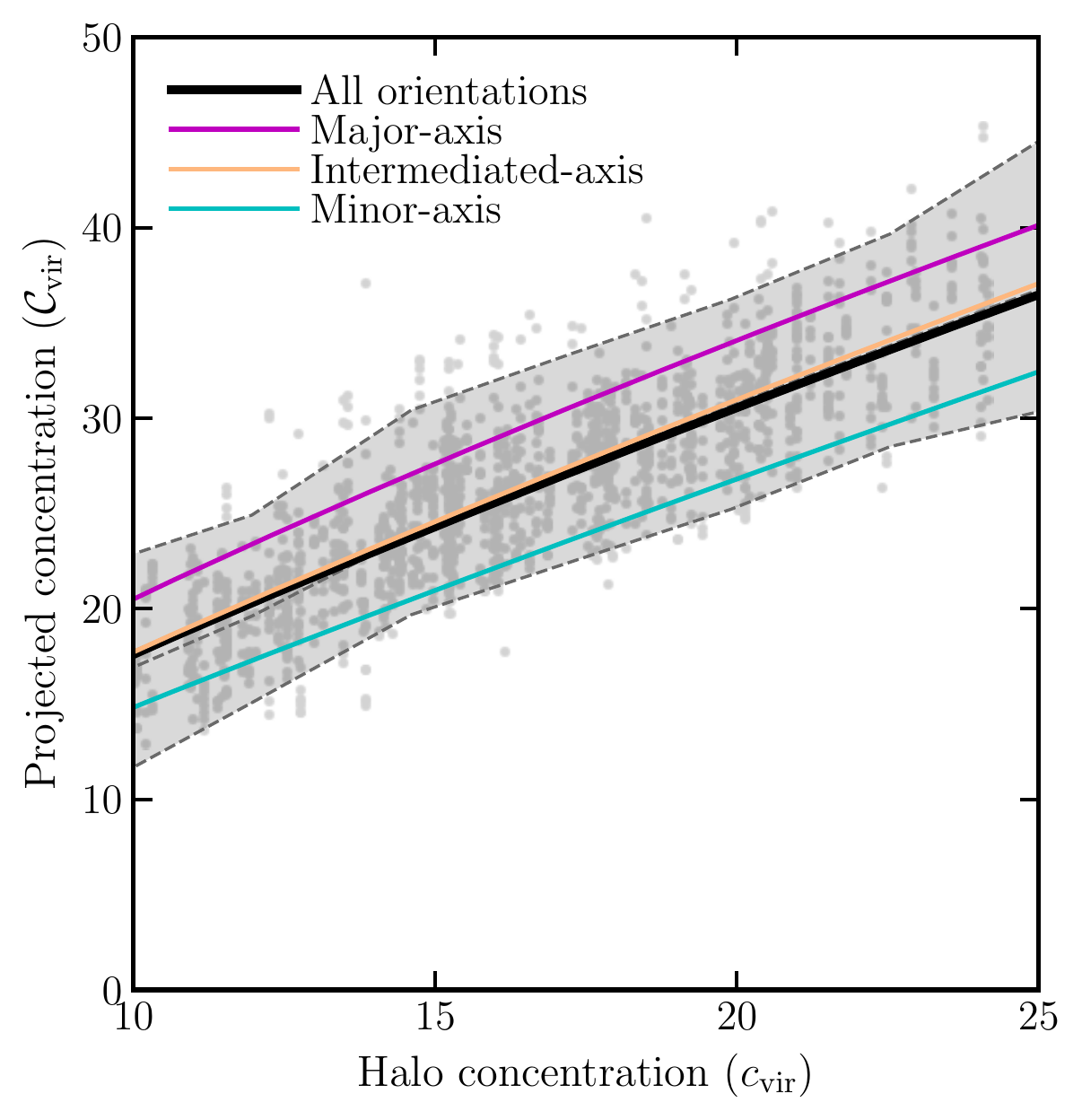}
    \caption{
        Relationship between the best-fit projected concentration using $\Sigma_{0.3}$ profiles, $\mathcal{C}_{\rm vir}$, and the standard three-dimensional halo concentration, $c_{\rm vir}$, determined for the same halos' best-fit Einasto profiles. The gray points correspond to 10 projected random orientations of all halos in our sample. The thick, solid black is a power-law fit to all of light-gray points while the dashed lines reflect the $2-\sigma$ scatter at fixed $c_{\rm vir}$. The colored solid lines are power-law fits to the projected fits along the major, intermediate, and minor axis of each halo. 
    }
    \label{fig:cvir-ceff}
\end{figure}

Another way to understand the utility of $\Sigma_{0.3}$ parameterization is compare its accuracy to the typical NFW approach at a fixed radius.  Figure~\ref{fig:fit.quality-innerDensity} shows the relative error (simulated vs. fit) of the two approaches at a projected radius $R=600\ \rm pc$, which corresponds to $\sim 0.2$ arcsec at $z=0.2$ as a function of halo virial mass.  In this exercise, we have viewed each simulated halo along 10 random orientations and used $\Sigma_{0.3}$ fits associated with each orientation compared with the implied spherical NFW fit in each case. Relative errors for the best-fit $\rho_{\rm NFW}$  and $\Sigma_{0.3}$ fits are shown in magenta and cyan, respectively. The solid curve shows the median as a function of $M_{\rm vir}$, while the bands encapsulates the $2\sigma$ range. We see that the relative error in NFW fits is larger at larger masses.  This reflects the fact that the $\Sigma_{0.3}$ fits show the most improvement for $R < R_{-1}$. At higher masses,  $600$ pc is ``deeper" within the core, so that the relative error coming from the NFW assumption increases.  

\subsection{The projected concentration parameter}
\label{sec:param.orient}
The spherically-averaged, three-dimensional density profile of a dark matter halo is often quantified by an NFW profile using two parameters: a halo mass and a concentration parameter \citep[e.g.][]{Bullock2001}. Similarly, we can quantify the circularly-averaged, projected density profile of a dark matter halo along any orientation with a {\em projected concentration}, defined as
\begin{align}
    \mathcal{C}_{\rm vir} := r_{\rm vir}/ R_{-1}
    \, .
    \label{eq:eff.conc}
\end{align}

Figure~\ref{fig:m1kpc-ceff} provides some physical insight on the meaning of these parameters. We plot the projected mass within 1 kpc, $\mathcal{M}(< R=1\ \rm kpc)$, as a function of halo mass.  For reference, 1 kpc has an angular size of $\sim 0.3$ arcsec at $z=0.2$. Each halo in our sample is viewed along 10 random orientations, which can be seen as vertical stacking at fixed virial mass at high masses. The color bar maps to the best-fit projected concentration $\mathcal{C}_{\rm vir}$ for each halo/orientation. Lower-density projections for the same halo have lower projected concentrations, and this provides a way to quantify the scatter associated with halo shape. As one might expect, there is a correlation between $\mathcal{M}(1\ \rm kpc)$ and $M_{\rm vir}$. For virial masses below $10^{10} M_{\odot}$, where $R_{-1} \lesssim 1$ kpc, the median relation scales as $\mathcal{M}(<1 {\rm kpc}) \sim M_{\rm vir}^{1/2}$. At higher masses, $R_{-1} \gtrsim 1$kpc, and the relationship becomes even flatter $\mathcal{M}(< 1 {\rm kpc}) \sim M_{\rm vir}^{1/3}$. The weak power-law dependence makes inferring the halo mass from a projected mas somewhat challenging; a 20\% error in surface density would map to a $\sim 40-60\%$ error in interpreted virial mass on average. The factor of $\sim 2$ scatter in projected mass at fixed virial mass is roughly captured by the scatter in projected concentration. Note that this level of variation at fixed virial mass is considerable; in the median, a factor of $2$ increase in projected mass at $1$ kpc is equivalent to a factor of $\sim 4$ ($8$) increase in $M_{\rm vir}$ at low (high) virial mass. 

The light gray points in Figure~\ref{fig:cvir-ceff} show the relationship between the projected concentration, $\mathcal{C_{\rm vir}}$, and the standard three-dimensional halo concentration, $c_{\rm vir}$, for individual halos. Each halo has a fixed $c_{\rm vir}$, but has 10 values of $\mathcal{C_{\rm vir}}$ derived from $\Sigma_{0.3}$ fits from 10 random orientations. We only include halos more massive than $M_{\rm vir} = 10^9\, M_\odot$ in this analysis in order to ensure accurate fits. The thick, solid black line shows a power-law fit to all the points, which provides a good representation of the average relation: $\mathcal{C_{\rm vir}} \propto c_{\rm vir}^{\delta}$. The best-fit function is found to be $\mathcal{C_{\rm vir}} = 2.3\, c_{\rm vir}^{0.87}$. The dashed lines show the $2\sigma$ distribution at fixed three-dimensional halo concentration, $c_{\rm vir}$. For reference, we also plot power-law fits for the same relation using only their major-, intermediate-, and minor-axis projections as pink, orange, and cyan lines, respectively. The intermediate-axis is roughly consistent with the average over all orientations, while the averages of the two other orientations track the $\sim 1.5\sigma$ range, suggesting that much of the variation at fixed $c_{\rm vir}$ comes from orientation effects.  

\begin{figure}
    \includegraphics[width=\columnwidth]{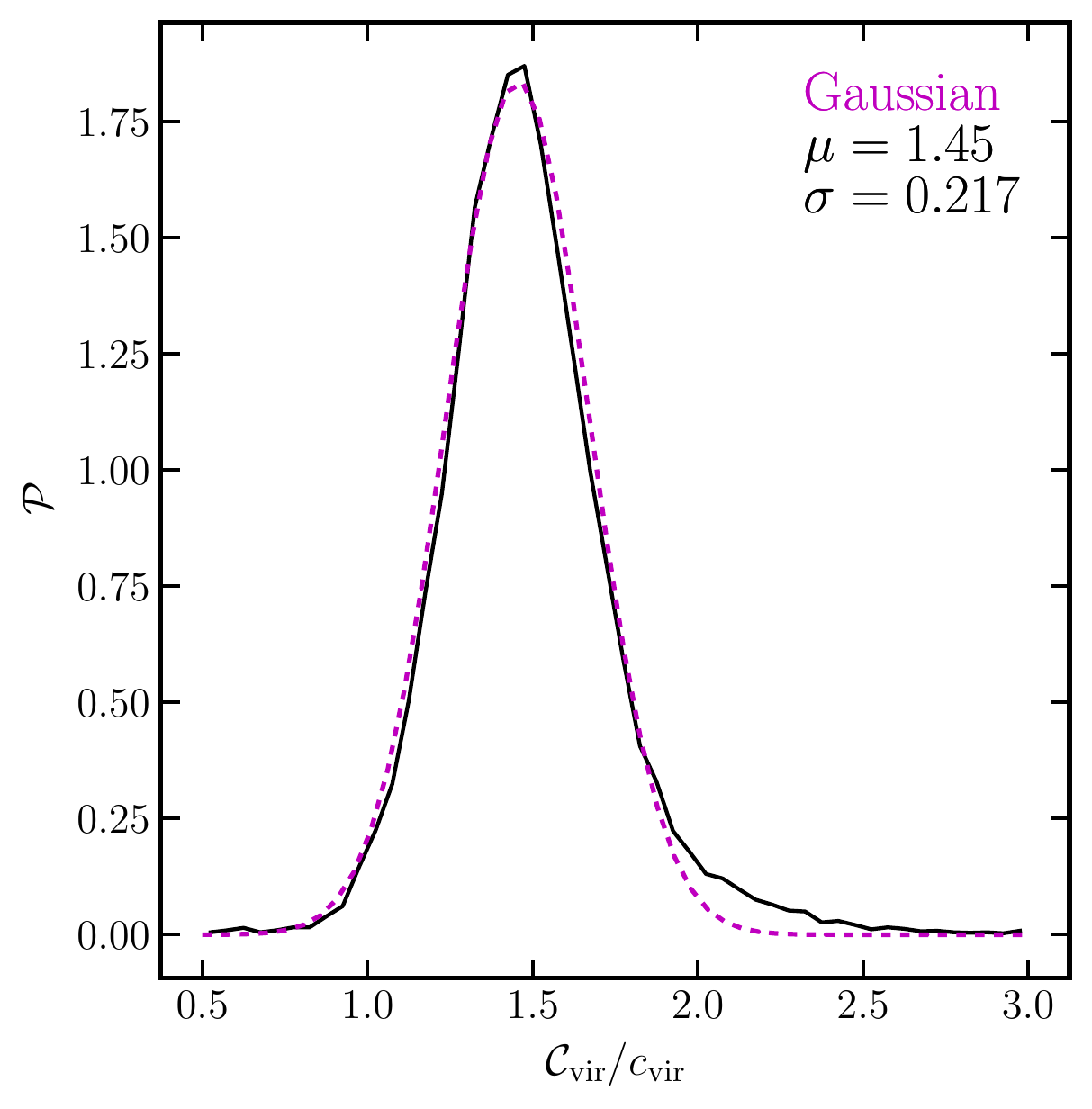}
    \caption{
    The probability distribution of the ratio between the projected concentration and the halo concentration $\mathcal{C}_{\rm vir}/c_{\rm vir}$ (black line), along with the best fit Gaussian fit (dashed magenta line). 
    }
    \label{fig:gaussian}
\end{figure}

Figure~\ref{fig:gaussian} plots the probability distribution (black curve) of $\mathcal{C}_{\rm vir}/c_{\rm vir}$ of all halos. The curve is nicely described by a Gaussian (dashed magenta) with mean $\mu = 1.45$ and standard deviation $\sigma = 0.22$. From this, we can infer that the parameters have a mean relation of $C_{\rm vir} \simeq 1.5\, c_{\rm vir}$ at this fixed redshift; a larger sample of halos, sampled in cosmological volumes, and studied at a wider range of redshifts, will be needed to explore these relationships more completely. 

Given the relationship  $\mathcal{C_{\rm vir}} \propto c_{\rm cvir}^{0.87}$ shown in Figure \ref{fig:cvir-ceff}, we have also explored an alternative to Figure \ref{fig:gaussian} that looks at the distribution of ratio $\mathcal{C_{\rm vir}}/c_{\rm vir}^{0.87}$. We find a similarly-shaped distribution, also well-described by a Gaussian, this with mean $\mu = 2.14$ and standard deviation $\sigma = 0.28$.  In this case the scatter relative to the mean is $\sigma / \mu \simeq 0.13$.

\begin{figure*}
    \centering
    \includegraphics[width=1.00\textwidth]{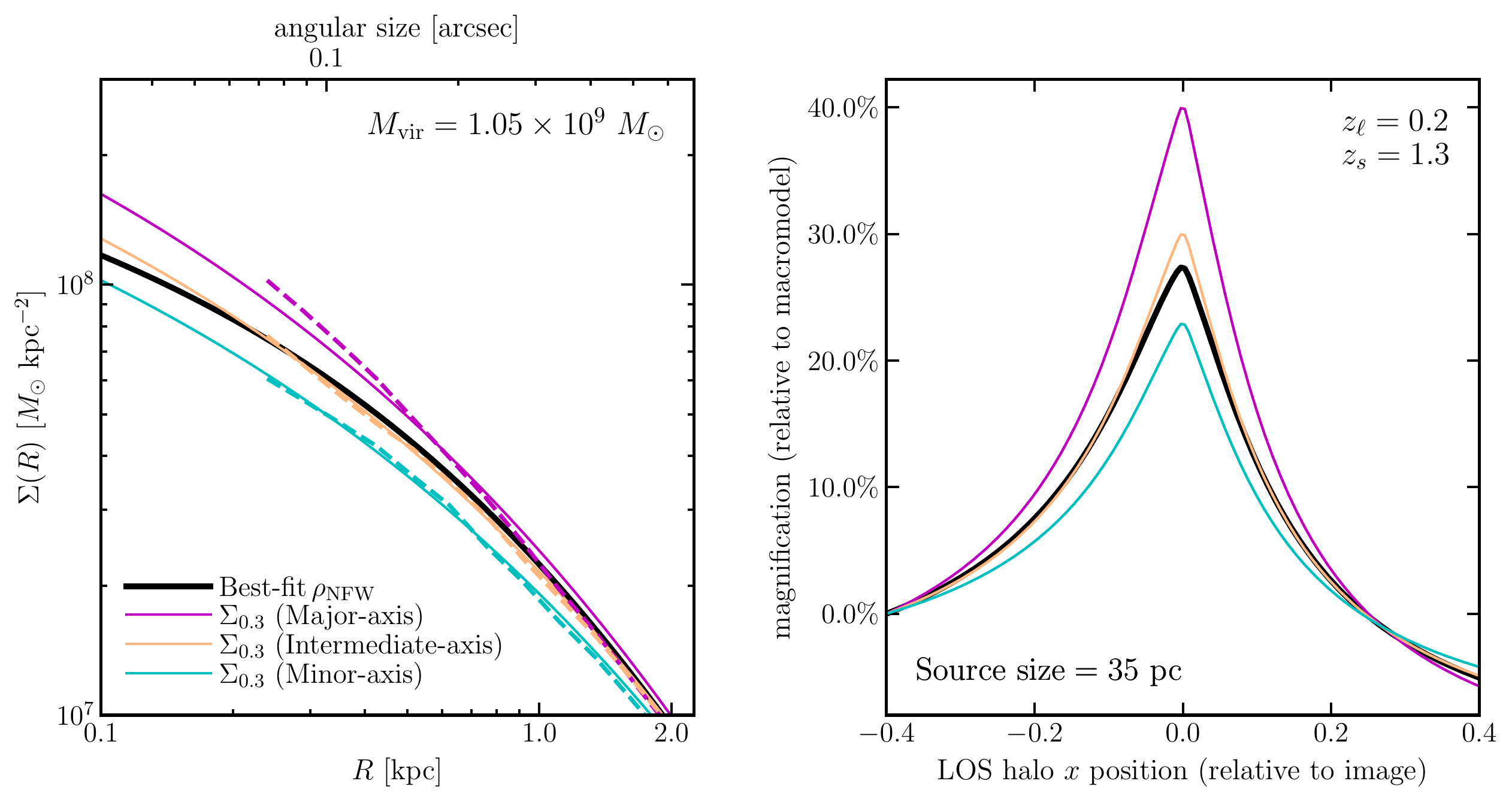}
    \caption{
    Comparison of lensing outcomes for $\Sigma_{0.3}$ fits and assumed spherical NFW fits for an example $M_{\rm vir} \simeq10^{9}\ M_{\odot}$ dark matter halo perturber. {\em Left}: Surface density profiles for the example halo. The thick, solid black curve is the implied surface density profile for the best-fit $\rho_{\rm NFW}$ for this halo. The dashed color curves show projected density profiles along the three primary density axes.  The respective $\Sigma_{0.3}$ fits are solid lines plotted in the same color as the dashed curves. {\em Right}: the relative change in magnification for an image relative to a smooth lens derived from the same analytic profiles on the left. We assume a source at redshift $z_{s}=1.3$ and lens at $z_{\ell}=0.2$. The horizontal axis shows the angular separation between the perturber halo and the image in arcseconds.  The $\Sigma_{0.3}$ profiles (colored curves) produce significant differences in relative magnification compared to the predicted NFW signal.  
    }
    \label{fig:mag.examp}
\end{figure*}

\section{Implications for gravitational lensing}
\label{sec:discussion}
Our primary motivation for characterizing the surface density profiles of dark matter halos is to provide a useful and accurate tool for gravitational lensing applications.  As shown in sections \ref{sec:deflection_potential}, \ref{sec:deflection_angle}, and \ref{sec:lensing_shear}, the $\Sigma_{\beta}$ profile given by Equation~\ref{eq:general.proj.profile} has an analytic deflection potential, deflection angle, and lensing shear. One example use case is in the search for substructure within multiply-imaged quasar strong-lens systems.  For example, the  magnification ratios of images compared to those of a smoothly-parameterized mass distributions from a macro model provides a way to discover and characterize invisible dark matter halo perturbers \citep{Mao1998,Metcalf2001,Dalal2002,Nierenberg2014,Nierenberg2017,Gilman2020a,Gilman2020b,Hsueh2020}. 

Figure~\ref{fig:mag.examp} shows an example of how the use of the $\Sigma_{0.3}$ profile gives different magnification predictions than an NFW profile fit for the same halo. For this calculation we use the open-source gravitational lensing software {\small LENSTRONOMY}\footnote{\href{https://github.com/sibirrer/lenstronomy}{https://github.com/sibirrer/lenstronomy}}, which performs the main lensing computations such as ray-tracing and magnifications \citep{Birrer2018,Birrer2021}, combined with the open-source {\small pyHalo}\footnote{\href{https://github.com/dangilman/pyHalo}{https://github.com/dangilman/pyHalo}}, which generates realizations for subhalos and LOS halos within a lens system \citep{Gilman2021}. The left side of Figure~\ref{fig:mag.examp} shows the projected surface density profile for an example $M \simeq 10^{9}\ M_{\odot}$ halo at $z=0.2$ along the three primary projections (colored lines). The lines are truncated at the convergence radius for this simulation.  Also shown is the implied surface-density profile for the best-fit (spherical) NFW profile for the same halo (solid black).  The three solid colored lines show the best-fit $\Sigma_{0.3}$ profile fits for each projection. Note that he horizontal axis on the bottom shows projected radius in physical units (kpc) and on the top we show the same quantity in angular size for a $z=0.2$ lens in seconds of arc. 

The right panel depicts the relative magnification cross-sections compared to a smooth model for the three-dimensional NFW fit (thick, solid black line) and the three best-fit $\Sigma_{0.3}$ models for each orientation. The colored lines on the right panel were calculated using the analytic fits shown in the same color on the left. We assume a Gaussian source light distribution with a FWHM of 35 parsecs (comparable to the size of the nuclear narrow-line region of a quasar; e.g., \citealt{Nierenberg2014}, \citealt{Muller-Sanchez2011}) for a source redshift $z_{\rm s} = 1.3$ and lens redshift $z_{\ell} = 0.2$.  This figure is analogous to Figure 2 in \citet{Gilman2020a}. The horizontal axis shows the perturber position relative to the image's position in arcseconds. The resulting change in magnification compared to a smooth-lens model is shown on the vertical axis. 

We see that when a halo is projected along the densest and least-densest axis, relative magnification changes by a factor of $\sim 2$. The best-fit NFW density profile (black solid curve) fails to match these two extremes. This level of difference could bias the inferred parameters of the perturbing halo systematically at the $\sim 50\%$ level.  For example, if we tried to interpret the major or minor axis orientations' magnification signals with NFW profiles of the same concentration, we would infer halos that are more or less massive than they actually are by factors of order unity.

\section{Summary}
\label{sec:conclusion}

We have studied the surface density structure dark matter halos using a suite of dark matter only zoom simulations  that contain halos of mass $M_{\rm vir} = 10^{7-11}\ M_{\odot}$.  We find that their cylindrically-averaged surface density profiles along any projection can be modeled as a function of projected radius $R$ using an easy-to-use, analytic profile, $\Sigma_{0.3}$,  defined in  Equation~\ref{eq:general.proj.profile} with $\beta = 0.3$.  The function has two free parameters: a scale radius $R_{-1}$ and a normalization $\Sigma_{-1}$. A primary motivation is to provide a profile that is more accurate than a projected spherical NFW profile and that has analytic lensing properties (\ref{sec:deflection_potential}, \ref{sec:deflection_angle}, and \ref{sec:lensing_shear}) that can be used in strong lensing studies for line-of-sight perturbing halos.   

The common approach to modeling small-halo perturbers is to assume spherical NFW profiles with properties drawn from mass-concentration relation predictions \citep[e.g.][]{Gilman2020b}.  As shown in Figures \ref{fig:surface.density.compare} and \ref{fig:fit.performance}, the  $\Sigma_{0.3}$ approach allows for significant improvement, with the same number of free parameters, especially at small projected radii, where the densities are highest and of most relevance for lensing perturbations.   An important by-product introduced here is the {\em projected concentration}, defined by the size of the halo and the projected scale radius,   $\mathcal{C}_{\rm vir} := r_{\rm vir}/ R_{-1}$. For any individual halo, the projected concentration can be higher or lower depending on the projected orientation. By fitting halos along an ensemble of orientations, the projected concentration provides a way to statistically account for asphericity while still utilizing a two-parameter foundation. The the projected concentration correlates with the standard three-dimensional concentration in a way that allows ease of use (Figures~\ref{fig:cvir-ceff} and~\ref{fig:gaussian}).   

In order to illustrate implications for gravitational lensing, we set up a mock-lensing analysis to explore the implied perturber magnification  with the $\Sigma_{0.3}$ model compared to the standard NFW approach (Figure \ref{fig:mag.examp}).  In this exercise, we examined the projected density structure of an example halo along three primary density axes and showed that the implied magnification perturbation as captured by $\Sigma_{0.3}$ fits for different orientations give as much as $\sim 50\%$ relative differences compared to the same halo's NFW fit at fixed mass. Flux ratio studies of lensing by halos draw populations of halos with halo to halo scatter in the mass-concentration relation. To studying the impact of the orientation observed in this work on a full dark matter inference would require full simulations of gravitational lenses. However we can see that the effect should in principle be similar to increasing the  ``effective scatter'' in the lensing mass-concentration relation \citep{Gilman2020a,Gilman2020b}. We leave a detailed analysis of this to a future work.   

One weakness of this analysis is that we have relied on zoom simulations, and not a fair cosmological sample. A next step in this analysis will be to use cosmological samples of simulated halos to provide statistically meaningful predictions for effective concentration distributions as a function of halo mass. We expect that the relative relationships we have found between the three-dimensional concentration and effective (projected) concentrations for our halos will be robust.  From the analysis done here, the average mapping obeys $\mathcal{C_{\rm vir}} = 2.3\, c_{\rm vir}^{0.87}$ (Figure \ref{fig:cvir-ceff}) with a Gaussian scatter of $\sigma \simeq 0.2$ at fixed $c_{\rm vir}$ for any individual halo. Future work will allow us to test these expectations in a fair cosmological context.

\section*{Acknowledgements}
We thank the anonymous referee for their helpful comments in improving the early version of this article.
Part of this research was carried out at the Jet Propulsion Laboratory, California Institute of Technology, under a contract with the National Aeronautics and Space Administration.
JSB was supported by the National Science Foundation (NSF) grant AST-1910965 and NASA grant 80NSSC22K0827. 
AL was supported by NASA grant 80NSSSC20K1469.
MBK acknowledges support from NSF CAREER award AST-1752913, NSF grants AST-1910346 and AST-2108962, NASA grant 80NSSC22K0827, and HST-AR-15809, HST-GO-15658, HST-GO-15901, HST-GO-15902, HST-AR-16159, HST-GO-16226, HST-GO-16686, HST-AR-170248, and HST-AR-17043.
LAM acknowledges partial support by the NASA Astrophysics Theory Program investigation 17-ATP17-120.
The analysis in this manuscript made extensive use of the python packages {\small NumPy} \citep{Van2011}, {\small SciPy} \citep{Oliphant2007}, {\small Matplotlib} \citep{Hunter2007}, and {\small COLOSSUS} \citep{Diemer2018}; We are thankful to the developers of these tools. This research has made all intensive use of NASA's Astrophysics Data System (\url{https://ui.adsabs.harvard.edu/}) and the arXiv eprint service (\url{http://arxiv.org}).

\section*{Data Availability}
The data supporting the plots within this article are available on reasonable request to the corresponding author. A public version of the {\footnotesize GIZMO} code is available at \url{http://www.tapir.caltech.edu/~phopkins/Site/GIZMO.html}. Additional data from the FIRE project, including simulation snapshots, initial conditions, and derived data products, are available at \url{https://fire.northwestern.edu/data/}.

\bibliographystyle{mnras}
\bibliography{references}

\appendix
\section{Lensing profile Properties}
\subsection{Projected mass distribution}
This paper discusses a surface density profile $\Sigma_{\beta}(R)$ for dark matter halos
described by Equation~\ref{eq:general.proj.profile}. The projected cumulative mass of the dark matter, $\mathcal{M}$, with a projected radius, $R$, for such a functional form is given by
\begin{align}
    \mathcal{M}(R) &= \int_{0}^{2\pi}\mathrm{d}{\phi} \int_{0}^{R}\mathrm{d}R'\, R'\, \Sigma_{\beta}(R') 
    \\
    &= 2\pi \int_{0}^{R}\mathrm{d}R'\, R'\, \Sigma_{\beta}(R') 
    \, ,
\end{align}
where the last line assumes spherical symmetry. The projected mass profile for the enclosed mass is
\begin{align}
    \mathcal{M}\left( R \right) = \frac{2\pi\Sigma_{-1}R_{-1}^{2}}{\beta} \exp\left({\frac{1 + 2\ln\beta}{\beta}}\right)\, \gamma\left[ \frac{2}{\beta},\, \frac{1}{\beta} \left( \frac{R}{R_{-1}} \right)^{\beta} \right] \, ,
        \label{eq:mass1}
\end{align}
where $\gamma(a,x)$ is the lower incomplete gamma function.

\subsection{Projected mass normalization}
 To determine the normalization relation, assuming circular symmetry, we integrate $R$ out to a virial radius, $r_{\Delta}$:
\begin{align}
    \mathcal{M}_{\rm eff} &= 2\pi \int_{0}^{r_{\Delta}}\mathrm{d}R'\, R'\, \Sigma_{\beta}(R') 
    \\ 
    \mathcal{M}_{\rm eff} &= 2\pi\, \Sigma_{-1} \int_{0}^{r_{\Delta}}\mathrm{d}R'\, R'\, \exp\left\{ -\frac{1}{\beta}\left[ \left(\frac{R'}{R_{-1}} \right)^{\beta} - 1 \right] \right\}\, .
\end{align}
Note that we have chosen $r_{\Delta}$, and then later $\mathcal{C}_{\Delta}$, to explicitly show that this works for any over-density definition used. We have also defined $\mathcal{M}(r_{\Delta}) \equiv \mathcal{M}_{\rm eff}$, since integrating out to the virial radius does not always ensure we recover the virial mass of the halo, i.e. $\mathcal{M}(r_{\Delta}) \neq M_{\rm vir}$, as the depth of the projection is set by a cylinder depth defined in the main text, 

. Moving forward, we can make the substitution $x = R/R_{-1}$, which leads us to the relation: 
\begin{align}
    \mathcal{M}_{\rm eff} &= 2\pi \Sigma_{-1} r_{\Delta}^{2}\, \mathcal{I}(\mathcal{C}_{\Delta}|\beta)\, ,
        \label{eq:mass.fit}
\end{align}
where
\begin{align}
    \mathcal{I}(\mathcal{C}_{\Delta}|\beta) &= \frac{1}{\mathcal{C}_{\Delta}^{2}} \int_{0}^{\mathcal{C}_{\Delta}}\mathrm{d}x\, x \exp\left\{ -\frac{1}{\beta}\left[ x^{\beta} - 1 \right] \right\}\, ,
    \label{eq:integral}
\end{align}
and the projected concentration is $\mathcal{C}_{\Delta}$.

\begin{figure}
    \centering
    \includegraphics[width=\columnwidth]{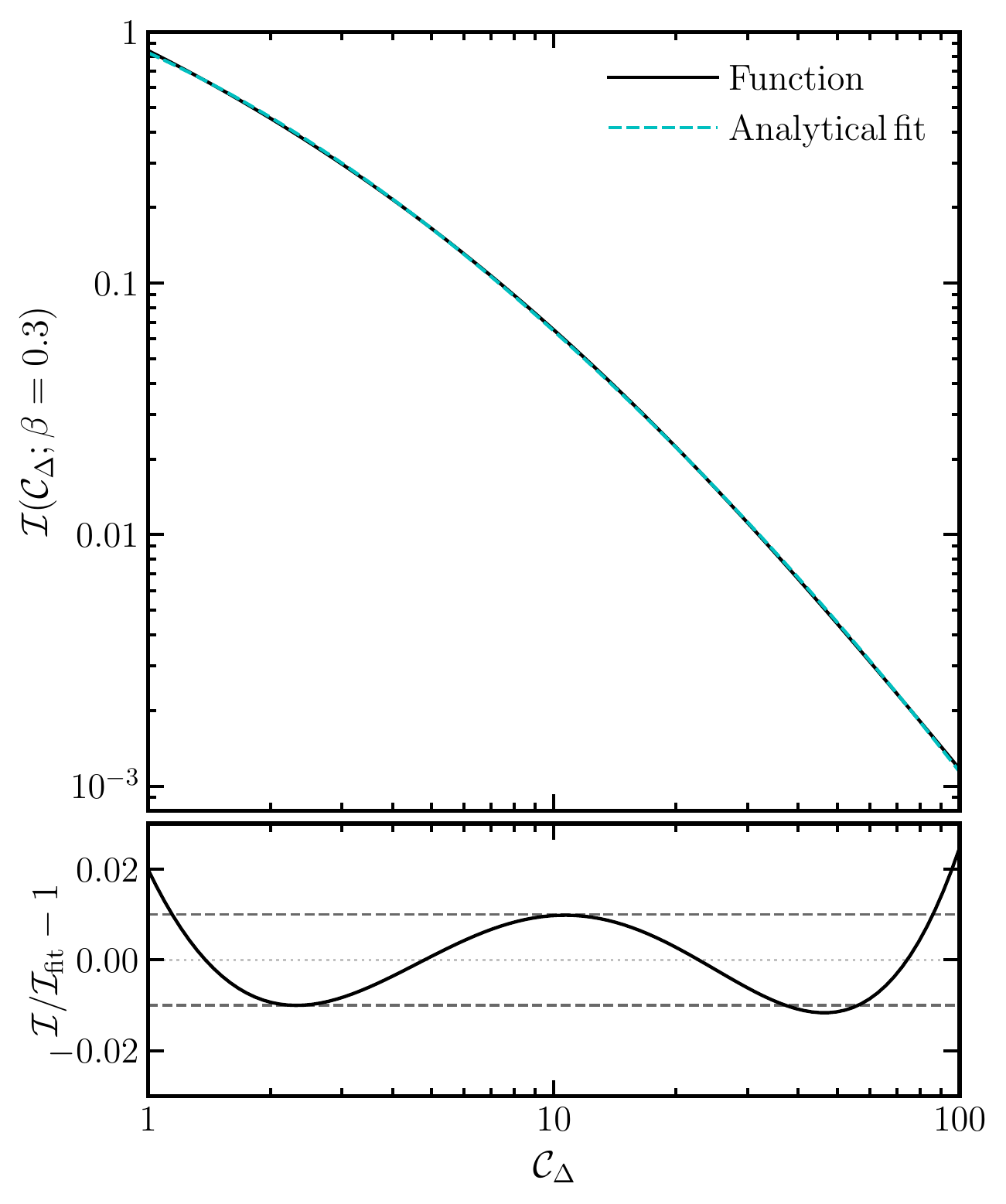}
    \caption{
    Fit to the integral function for $\beta = 0.3$ as a function of the projected concentration. The top panel plots Equation~\protect\ref{eq:integral} as the black line while the dashed cyan curve is the resulting fit using Equation~\protect\ref{eq:integral.fit}. The bottom panel plots the fit quality. The fit is accurate to 2\% and better for a large range of projected concentrations.
    }
    \label{fig:int.fit}
\end{figure}

The integral $\mathcal{I}$ cannot be solved analytically, but can be numerically integrated for a given value of $\mathcal{C}_{\Delta}$ and $\beta$. We provide a generalized fitting fitting function:
\begin{align} 
    \mathcal{I} = c_{0} \exp\left[ -n \left(\frac{\mathcal{C}_{\Delta} + c_{1}}{c_{2}}\right)^{\frac{1}{n}} \right]\, .
    \label{eq:integral.fit}
\end{align}
Once fitted with $\beta = 0.3$, the best-fit parameters $c_{0} = 5756$, $c_{1} = 0.628$, $c_{2} = 0.438$, $n=7.41$. The results are presented in Figure~\ref{fig:int.fit},  where the black line is the above integral with a fixed $\beta = 0.3$ as a function of the projected concentration, while the red-dashed line is the resulting fit. The bottom panel plots the residuals between the integral and the analytical fit. The best fit parameters is accurate to $\sim$ 2\% and better for ranges of $\mathcal{C}_{\Delta} \in [1,100]$. While not shown here, Equation~\ref{eq:integral.fit} also fits accurately for shapes with $\beta \in [0.01,0.40]$, which can be applicable for halos with differing shapes. With above results, we arrive with the final form of the profile normalization as a function of mass and projected concentration:
\begin{align}
    \Sigma_{-1}
    &= \frac{\mathcal{M}_{\rm eff}}{2\pi\, r_{\Delta}^{2}\, \mathcal{I}(\mathcal{C}_{\Delta} ; \beta)} 
    \, .
\end{align}
Note that this normalization can be connected to the virial definition of the halo by imposing an established $\mathcal{M}_{\rm eff}$--$M_{\rm vir}$ relation.

\subsection{Unit-convergence radius}
The surface mass-density of a dark matter halo with a spherical $\Sigma_{\beta}$ profile (Equation~\ref{eq:general.proj.profile}) expressed in units of the critical surface density yields the dimensionless convergence profile:
\begin{align}
    \kappa_{\beta}(R) = \Sigma_{\beta}(R)/\Sigma_{\rm c} = \kappa_{-1} \exp\left\{ -\frac{1}{\beta} \left[ \left( \frac{R}{R_{-1}}\right)^{\beta} - 1 \right]\right\}
    \label{eq:convergence.profile}
\end{align}
with $\kappa_{-1} := \Sigma_{-1}/\Sigma_{\rm c}$  such that $\Sigma_{-1} := \Sigma(R_{-1})$ and
 $\Sigma_{\rm c}$ is the critical surface density for lensing,
\begin{align}
    \Sigma_{\rm c} := \frac{c^{2}D_{s}}{4\pi D_{\ell} D_{\ell s}} \, ,
\end{align}
such that $D_{s},\, D_{\ell}$ and $D_{\ell s}$ are the angular diameter distances between the observer from the source, the observer from the lens, and the lens from the source. The unit-convergence radius, $R_{\kappa}$, at which the convergence $\kappa_{\beta}(R_{\kappa}) = 1$, is simply
\begin{align}
    R_{\kappa} = R_{-1} \left[ \beta\log\left(\kappa_{-1}\right) + 1 \right]^{1/\beta}
    \, .
\end{align}

\subsection{Deflection potential}
\label{sec:deflection_potential}
A crucial component of modern lensing software is to analytically implement the lensing potential (and deflection angle) for structure in projection. Our projected mass profile is conveniently in a form similar to that of a two-dimensional \cite{Sersic1968} light profile. Moreover, the lens properties for a \cite{Sersic1968} profile are already derived in \cite{Cardone2004}, making it a straightforward procedure to map our parameters to those results. For completeness, we re-derive the lensing properties in a similar manner to \cite{Cardone2004}.

With the $\Sigma_{\beta}$ profile (Equation~\ref{eq:general.proj.profile}), it is convenient to introduce a dimensionless variable $y := \left(R/R_{-1}\right)^{\beta}$. The two-dimensional lens potential, $\psi(R)$, can be determined by solving the two-dimensional Poisson equation (e.g. \citealt{Schneider1992}):
\begin{align}
    \nabla^{2}\psi(R) = 2\, \kappa_{\beta}(R)\, ,
\end{align}
where $\kappa_{\beta}(R)$ is the convergence profile for our model defined previously in Equation~\ref{eq:convergence.profile}. Expanding out the two-dimensional Laplacian on the left-hand side to polar coordinates, the following change of variables transforms the differential equation as
\begin{align}
    y^{2\left(1-\frac{1}{\beta}\right)}\frac{\mathrm{d}^{2}\psi}{\mathrm{d}y^{2}} + y^{\left(1-\frac{2}{\beta}\right)}\frac{\mathrm{d}\psi}{\mathrm{d}y} 
    = \frac{2\, R_{-1}^{2}\, \kappa_{-1}}{\beta^{2}}  \exp\left[ -\frac{1}{\beta} (y-1)\right]\, .
\end{align}
The above equation has the solution
\begin{align}
    \psi(y) = \psi_{-1} y^{\frac{2}{\beta}}\,
    _{2}F_{2}\left( \frac{2}{\beta},\frac{2}{\beta} ; 1+\frac{2}{\beta}, 1+\frac{2}{\beta}; \frac{y}{\beta} \right)\, ,
\end{align}
where we have used the conditions that $\psi \rightarrow 0$ as $y \rightarrow 0$ or $\infty$, $_{p}F_{q}\left( a_{1}, \dots, a_{p} ; b_{1}, \dots, b_{p} ; z\right)$ is the generalized hypergeometric function \citep{Gradshteyn1980}, and
\begin{align}
    \psi_{-1}
    &:=
    \frac{R_{-1}^{2} \kappa_{-1}}{2\, e^{1/\beta}}
    \, 
\end{align}
is the value of the potential for $y = 1$.

\subsection{Scaled deflection angle}
\label{sec:deflection_angle}
Our lens profile model imposes circular symmetry. Therefore, the deflection angle is purely radial and will have a magnitude $\alpha$ given by $\mathrm{d}\psi/\mathrm{d}R$. The complete differentiation of the potential with respect to $y$ gives
\begin{align}
    \alpha(y) 
    &= 2\, \alpha_{-1}\, y^{-\frac{1}{\beta}}\,
    \left[ \Gamma\left(1 + \frac{2}{\beta}\right)- \frac{2}{\beta}\, \Gamma\left[\frac{2}{\beta},-\frac{y}{\beta}\right] \right]\, ,
\end{align}
where $\Gamma(x,a)$ is the incomplete Gamma function, $\Gamma(a)$ is the standard Gamma function, and
\begin{align}
    \alpha_{-1}
    &:=
    \frac{\left(- \frac{1}{\beta}\right)^{-\frac{2}{\beta}}}{e^{1/\beta}\, \beta }\, R_{-1}\,  \kappa_{-1}
    \, .
\end{align}

\subsection{Lensing shear}
\label{sec:lensing_shear}
The radial dependence of a spherically symmetric system can be written as
\begin{align}
    \gamma_{\ell}(y) = \frac{\overline{\Sigma}(y) - \Sigma(y)}{\Sigma_{\rm c}} \, ,
\end{align}
where
\begin{align}
    \overline{\Sigma}(y)
    = 
    \frac{2}{y^{2}} \int_{0}^{y} \mathrm{d}y'\, y'\, \Sigma(y')
\end{align}
is the mean surface mass density inside the dimensionless radius $y$. For our lensing model, if we define $\tilde{y} := y^{\beta}/\beta$, the mean surface mass density is then
\begin{align}
    \overline{\Sigma}_{\beta}(\tilde{y})
    = 
    2\, \Sigma_{-1} \exp\left( \frac{1 - \beta\ln\beta}{\beta} \right) \frac{1}{\tilde{y}^{2/\beta}}
    \, 
    \gamma\left[\frac{2}{\beta},\, \tilde{y}\right]
    \, ,
\end{align}
which gives us the shearing quantity in its compact form:
\begin{align}
    \gamma_{\ell}(y) = \kappa_{-1}
    \left\{ \exp\left( \frac{1 - \beta\ln\beta}{\beta} \right) \frac{2}{\tilde{y}^{2/\beta}}
    \, 
    \gamma\left[\frac{2}{\beta},\, \tilde{y}\right]
    - \exp\left(-\tilde{y}+1\right) \right\}
    \, .
\end{align}

\section{The impact of projection depth}
\label{sec:projection.depth}
\begin{figure}
    \includegraphics[width=\columnwidth]{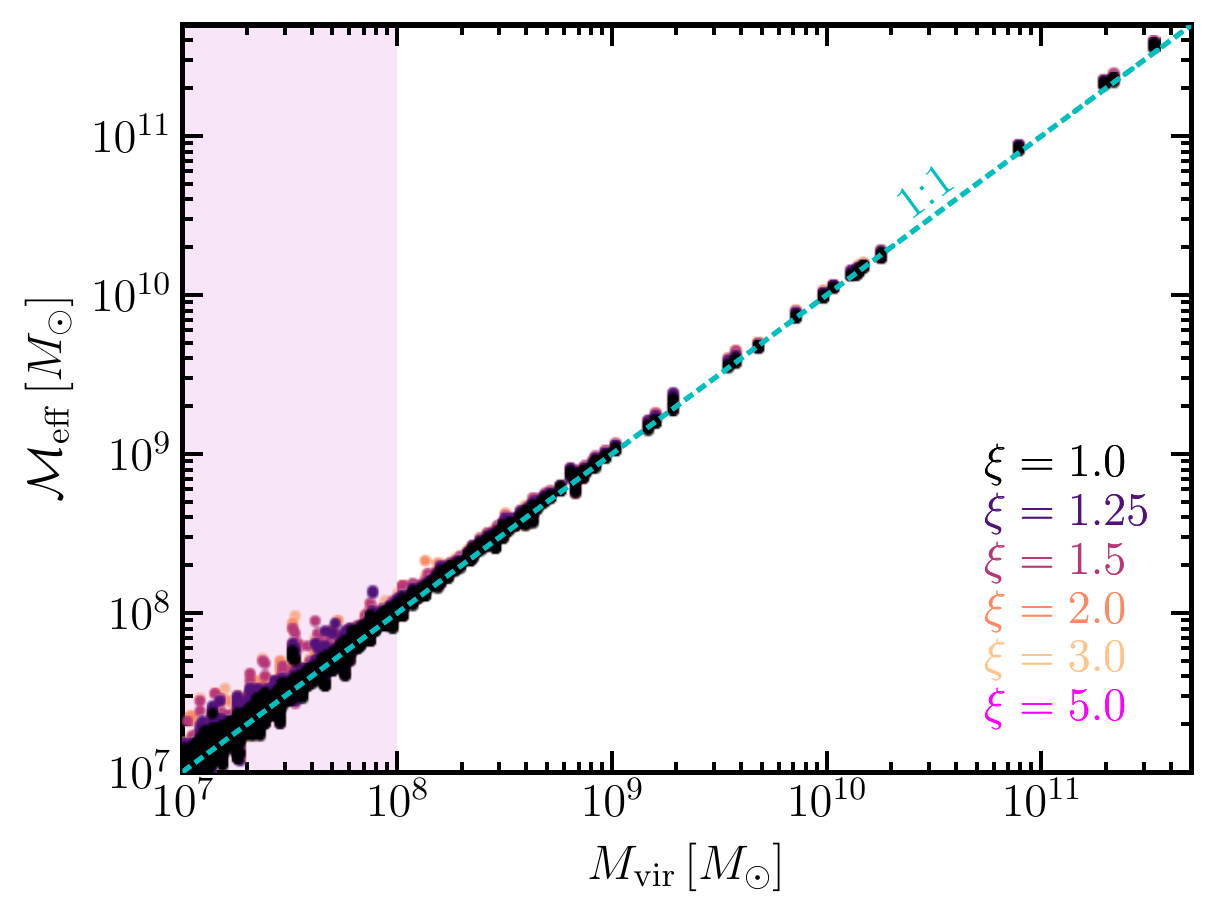}
    \includegraphics[width=\columnwidth]{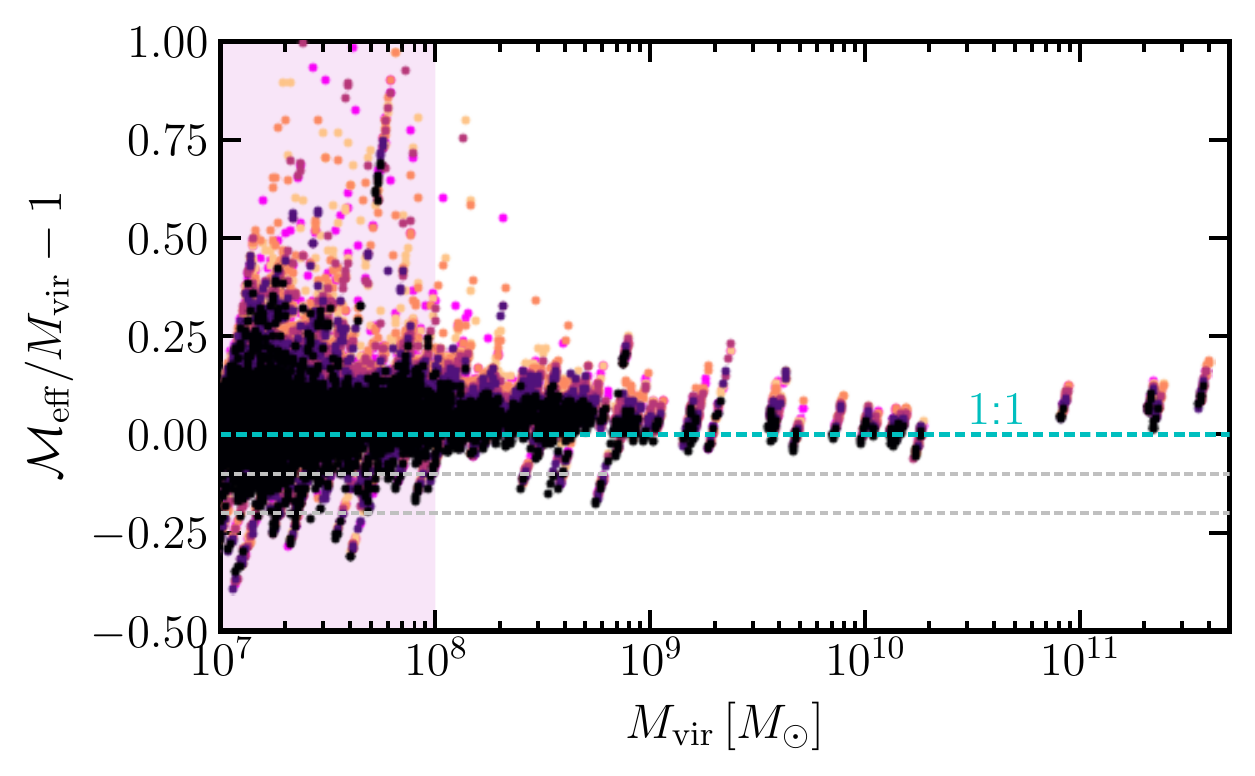}
    \caption{
    The virial mass plotted against the effective (projected) mass in multiple depths of projections for our chosen cylinder geometry: $\mathcal{L} = 2 \xi\, r_{\rm vir}$. The color codes map to the value of $\xi$ as shown. Note that the cylinder radius is fixed to be the virial radius. From the top plot, we see that the two masses are roughly be same. The cyan curve is a one-to-one relationship. The bottom plot shows residuals.  The shaded pink band indicates the region where the convergence radius of each halo is larger than $R_{-1}$.
    }
    \label{fig:Meff.vs.mvir}
\end{figure}

A major component within the analysis done in this paper is how the mass is collected within a given a dark matter halo along the projection. Here, we treat the line-of-sight projection as a face-on-viewed cylinder, with a radius and length, $\mathcal{R}$ and $\mathcal{L}$. The total mass within this volume is denoted at the {\em effective mass}, $\mathcal{M}_{\rm eff}$, which is assumed to relate to $M_{\rm vir}$ in some way. In the main text, we chose the radius to be the size of the halo, $\mathcal{R}=r_{\rm vir}$, while we chose the cylinder length to be proportional to the  halo virial radius $\mathcal{L} = 2 \xi\, r_{\rm vir}$. The chosen value $\xi$ plays a small part in inferring the effective mass. Indeed, as Figure~\ref{fig:Meff.vs.mvir} shows that the effective mass, based on our chosen cylinder volume, and the halo virial mass are roughly one-to-one, i.e., $\mathcal{M}_{\rm eff} \simeq M_{\rm vir}$ for most chosen values of $\xi$. Notably, as $\xi$ increases, the scatter (which has reduced Poisson noise) for the lower-mass halos increases. 

\begin{figure}
    \includegraphics[width=\columnwidth]{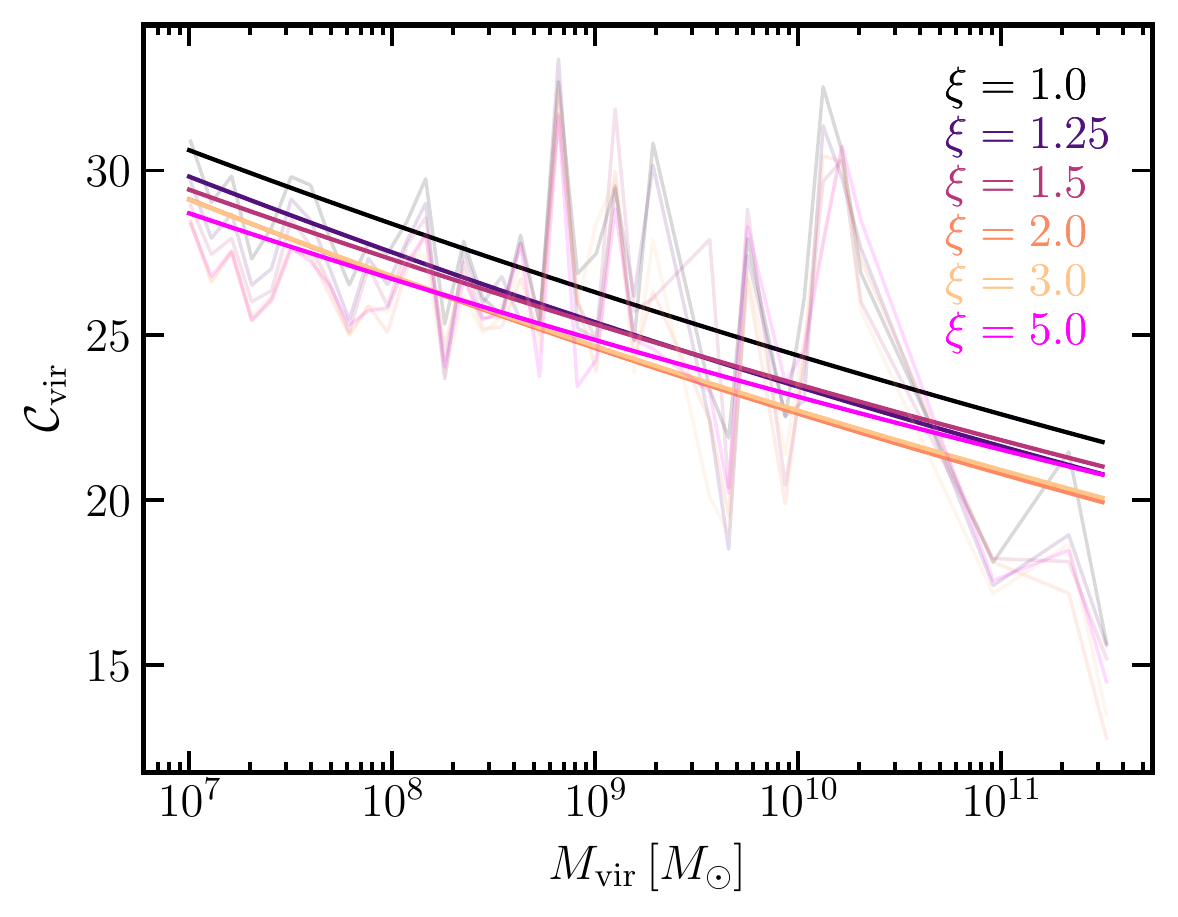}
    \caption{
    Same $\xi$ values from the previous figure, now plotting the implied projected concentration-mass relations. No changes of the normalization are present.
    }
    \label{fig:Cvir.vs.mvir}
\end{figure}

While the effective mass is consistent for different values of $\xi$, that is only one part of the complete story, where fortunately, the the inferred projected concentration is not impacted that much either. Figure~\ref{fig:Cvir.vs.mvir} shows very little variance in the normalization of the mass function for the corresponding values of $\xi$ previously shown in Figure~\ref{fig:Meff.vs.mvir}. These projected concentration functions are compiled using our sample of zoom simulations, which has varying initial conditions, under-dense regions, as compared to fully realized cosmological boxes, and a relation riddled Poisson noise; these results are to not to be taken as legitimate statistical relations and are plotted for demonstration purposes. The faded-solid lines depict the bin values of $\mathcal{C}_{\rm vir}$ while the solid lines are power-law fits to these curves, as we want to emphasize the normalization values. As we increase to larger values of $\xi$, we see the normalizations remains relatively the same with slight adjustments of the slopes. 

From both figures, it was decided to use a value of $\xi = 1.5$ as this is where the projected concentration and mass relation converges while also minimizing the scatter down at the low-mass range for the effective mass and halo mass relation (which otherwise large for $\xi = 2$).

\section{Logarithmic slope of surface density profile}
\label{sec:log.slopes}
Here, we demonstrate that the behavior of the logarithmic slope for the projected density profile is captured by a radially-dependent power law profile that is simple in form:
\begin{align}
     \frac{\mathrm{d}\log{\Sigma_{\beta}}}{\mathrm{d}\log{R}}
    =
    - \Bigg( \frac{R}{R_{-1}} \Bigg)^{\beta} 
    \, .
\end{align}
Here, $R_{-1}$ is the radius where the logarithmic slope of the surface density is equal to $-1$ and $\beta$ is the shape parameters of that effectively tailors itself to any dark matter halo.

\begin{figure*}
    \centering
    \includegraphics[width=1.0\textwidth]{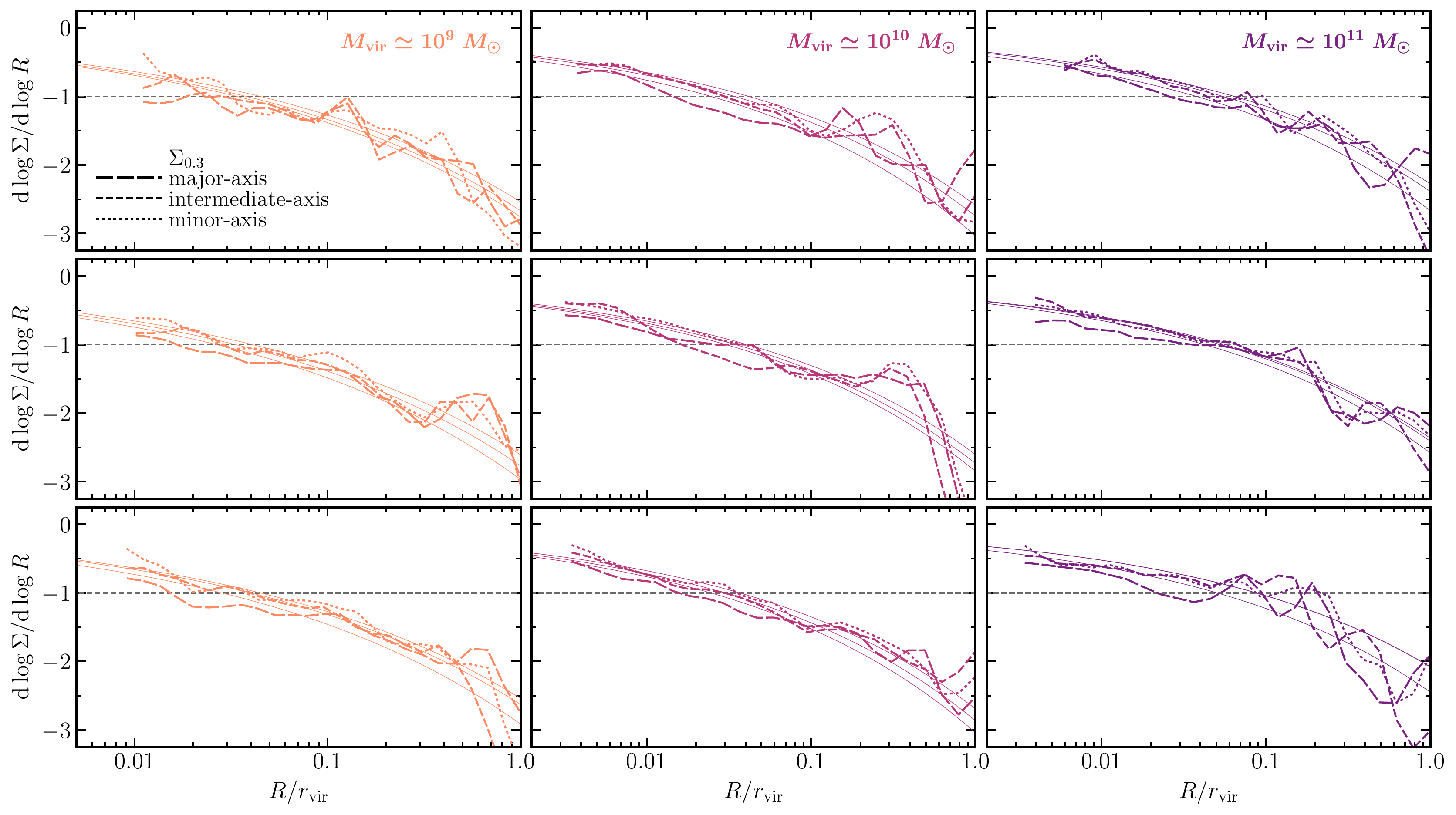}
    \caption{
        The logarithmic slope of the surface-mass density profiles, $\Sigma(R)$, along three density axis. Presented are several example halos for mass bins of $M_{\rm vir} \simeq 10^{11}$ (right-most panel), $10^{10}$ (center panel), and $10^{9}\ M_{\odot}$ (left-most panel) for demonstration purposes. The three density axes for a single halo are is plotted within a single panel. A power-law radial dependence of the slope seems to fit the different density axes of our simulations very well. Our predictive model, $\Sigma_{\beta}$, introduced in Equations~\protect\ref{eq:proj.profile:slope} and \protect\ref{eq:general.proj.profile}, are shown as thin-solid lines for the best-fit parameters to $\Sigma_{\beta}$ with fixed $\beta = 0.3$.
    }
    \label{fig:log.slope}
\end{figure*}

This is shown in Figure~\ref{fig:log.slope}, where we selected three representative halos within three of the higher-mass bins: $10^{9},\ 10^{10},$ and $10^{11}\ M_{\odot}$ in the left-most, center, and right-most panel, respectively. The logarithmic slopes in each mass bin follows a similar shape out to $r_{\rm vir}$ along each density-axes projection, minus the sporadic-like behavior seen in the outer region. It is until the profiles approaches the inner-region where we begin to see differences between all three profiles. A clear deviation, found for all of these masses, between each density axes is seen around the region at which the log-slope is equal to $-1$, which is marked by the dotted-gray horizontal line. Another important feature here is how each profile does not converge to some finite value. This would imply that the actual two-dimensional profile does not centrally converge to some power law. 

The thinner solid lines in each panel of Figure~\ref{fig:log.slope} depicts Equation~\ref{eq:proj.profile:slope}. The lines are determined from fitting the integrated profile of $\Sigma_{\beta}$ to the local surface density profiles with a fixed shape $\beta = 0.3$. The power-law profile is shown to reproduce the radial dependence fairly well within the three different density axes. Not that by eliminating the dependence of the shape parameter, $\beta$, for a given halo, a given orientation is parametrized by the projected scale radius, $R_{-1}$, which in-turn captures the variation and scatter expected from halo-to-halo.

\bsp
\label{lastpage}
\end{document}